\documentclass[aps,prl,superscriptaddress,twocolumn,longbibliography]{revtex4-1}
\usepackage{bbm}
\usepackage{graphicx}
\usepackage{dcolumn}
\usepackage{bm}
\usepackage{subfigure}
\usepackage{amsmath}
\usepackage{feynmf}
\usepackage{hyperref}

\usepackage{attachfile}

\newcommand{\bk}{\boldsymbol k}

\newcommand{\bq}{\boldsymbol q}

\newcommand{\bd}{\boldsymbol d}
\newcommand{\ba}{\boldsymbol a}
\newcommand{\bb}{\boldsymbol b}

\usepackage{times}

\begin{document}
\title{Sublattice-sensitive Majorana Modes}

\author{Di Zhu}
\altaffiliation{These authors contributed equally to this work.}
\affiliation{School of Physics, Sun Yat-Sen University, Guangzhou 510275, China}

\author{Bo-Xuan Li}
\altaffiliation{These authors contributed equally to this work.}
\affiliation{Beijing National Laboratory for Condensed Matter Physics and Institute of Physics,
Chinese Academy of Sciences, Beijing 100190, China}
\affiliation{University of Chinese Academy of Sciences, Beijing 100049, China}

\author{Zhongbo Yan}
\email{yanzhb5@mail.sysu.edu.cn}
\affiliation{School of Physics, Sun Yat-Sen University, Guangzhou 510275, China}

\date{\today}

\date{\today}

\begin{abstract}
For two- and three-dimensional topological insulators whose unit cells
consist of multiple sublattices, the boundary terminating at
which type of sublattice can affect the time-reversal invariant momentum at which the Dirac points of helical boundary
states are located.
Through a general theory and a representative model,
we reveal that this interesting property allows the realization of Majorana modes at sublattice domain
walls forming on the boundary when the boundary Dirac points of the topological insulator are gapped
 by an unconventional superconductor in proximity. Intriguingly, we find that the sensitive
sublattice-dependence of the Majorana modes allows their positions to be precisely manipulated
by locally controlling the terminating sublattices or boundary potential.
Our work reveals that the common sublattice degrees of freedom in materials open a new route to
realize and manipulate Majorana modes.
\end{abstract}

\maketitle

As a class of topological excitations,
Majorana modes in topological superconductors (TSCs)
have attracted tremendous research enthusiasm since a connection
to fault-tolerant quantum computation was built~\cite{Kitaev2003,nayak2008review}. On the road
to the final application in quantum computation, it is widely believed that
a milestone will be the implementation of braiding
Majorana zero modes (MZMs)~\cite{Sarma2015}, a type of bound-state Majorana modes.
Historically, as MZMs was initially revealed to appear in the vortex cores of
two dimensional chiral $p$-wave superconductors in the topological regime~\cite{read2000},
the initial scenario for braiding MZMs is based on the natural idea
of moving and exchanging vortices~\cite{Ivanov2001}. Later, theorists showed that
the braiding process can also be carried out in networks of one-dimensional
TSC wires~\cite{Alicea2011brading,Aasen2016}. Despite being viewed as two most promising
routes,  an experimental realization of either one of them remains elusive till date.
On one side, although steady and remarkable progress has been witnessed in the pursuit of
MZMs in platforms ranging from semiconductor
nanowires~\cite{Mourik2012MZM,das2012zero,deng2012anomalous,finck2013,rokhinson2012fractional,Deng2016mzm,
Albrecht2016,Chen2017mzm,Zhang2017nanowire,Fornieri2019,Ren2019mzm} and magnetic atom chains~\cite{Nadj2014MZM,Jeon2017,Kim2018mzm} to superconducting topological insulators~\cite{Xu2015MZM,Sun2016Majorana,Jack2019observation} and
iron-based superconductors~\cite{zhang2018iron,wang2018evidence,Liu2018MZM}, a decisive confirmation of MZMs in experiments
has not been achieved. On the other hand, both scenarios
require some levels of controllability
on the positions of MZMs, however, manipulating vortex-core or wire-end
MZMs in a highly controllable way itself is also rather challenging in experiments.

In the past few years, the birth of the concept named higher-order
TSCs provides new perspectives for both the
implementation and manipulation of both MZMs and other propagating Majorana modes~\cite{Benalcazar2017,Schindler2018,Song2017,Langbehn2017,Shapourian2018SOTSC,Khalaf2018,Geier2018,Zhu2018hosc,Yan2018hosc,
Wang2018hosc,Liu2018hosc,Wangyuxuan2018hosc,Hsu2018hosc,Wuzhigang2019hosc,
Volpez2019SOTSC,Zhang2019hinge,Gray2019helical,Zhu2019mixed,Peng2019hinge,Zhang2019hoscb,Yan2019hosca,Yan2019hoscb,Ghorashi2019,Bultinck2019,Zhang2020SOTSC,Ahn2020hosc,
Hsu2020hosc,Wu2020SOTSC,Pan2019SOTSC,Franca2019SOTSC,Majid2020hosca,Majid2020hoscb,Bitan2020hosc,Laubscher2020mcm,Zhang2020SOTSCb,Vu2020hotsc,wu2020boundaryobstructedb,
Wu2021fese,Pahomi2020,Tiwari2020chiral,Ikegaya2021mcm,Majid2021vortex,Li2021bts,Wu2021multiorder,Li2021mcm,Fu2021cmhm,Plekhanov2021,Majid2021surface,
Ghosh2021hierarchy,Luo2021hosc,chew2021higher,tan2021edge,jahin2021higher,li2021higher,scammell2021intrinsic}.
A unique characteristic
of higher-order TSCs is that the concomitant Majorana modes have
a codimension ($d_{c}$) larger than one and their locations in real space
depend on the boundary geometry, which is fundamentally distinct to
conventional TSCs (also dubbed first-order TSCs)
protected by internal symmetries only~\cite{Chiu2015RMP}, where the Majorana modes have $d_{c}=1$
and their locations do not rely on the boundary geometry as they appear everywhere on the whole boundary.
Because of the freedom on the boundary,  the positions of Majorana modes in
a higher-order TSC are in principle allowed to move if the
Majorana modes are not pinned by any crystalline symmetry~\cite{Langbehn2017,Geier2018}.
Indeed, previous works have shown that the positions of MZMs in two-dimensional
second-order TSCs can be tuned by rotating the orientation of magnetic field~\cite{Zhu2018hosc,Pahomi2020,Liu2021manipulation}
or changing the boundary potential via electrical gating~\cite{Zhang2020SOTSCb,Wu2021fese},
accordingly opening new routes to manipulate and braid MZMs~\cite{Zhang2020SOTSCa,Lapa2021braiding,Pan2021braiding}.
In this work, we reveal that the sublattice degrees of freedom
commonly appearing in materials admit a new intriguing scheme
for the realization and manipulation of Majorana modes with $d_{c}=2$.
Remarkably, this scheme can be applied to systems both with and without time-reversal symmetry (TRS),
and allows  the positions of Majorana modes to be precisely manipulated.

As putting first-order
topological insulators  in proximity to unconventional superconductors
can provide a natural realization of second-order TSCs~\cite{Yan2018hosc,Wang2018hosc,Liu2018hosc},
throughout this work we focus on this class of platforms to illustrate our theory.
Accordingly, the physics can be roughly descried as follows.
For a $d$-dimensional first-order topological insulator with $d>1$, while the appearance of helical
states does not depend on the terminating sublattice type on the boundary~\cite{hasan2010,qi2011},
a fact, interesting but having attracted little attention, is that
the terminating sublattice type can affect the time-reversal invariant momentum (TRIM)
at which the  Dirac points of helical boundary states are located. On the other hand, it is known that the
boundary Dirac points of a topological insulator can be gapped by the Dirac mass induced
by the superconductor in proximity~\cite{fu2008,Fu2009qshe}.
Notably, if the superconducting pairing is momentum-dependent,
both the magnitude and sign of the superconductivity-induced Dirac mass
depend on the location of the boundary Dirac point. This indicates that the terminating sublattice type
can directly affect the formation as well as the locations of
domain walls binding Majorana modes. Below we first formulate the
general theory from a boundary perspective, and then consider
a two-dimensional topological insulator with honeycomb lattice and proximity-induced
extended s-wave superconductivity
to demonstrate the physics.

{\it General theory from a boundary perspective.---}
Within the mean-field framework, a superconducting
system can be described by a corresponding
Bogoliubov de-Gennes (BdG) Hamiltonian of the form
$H=\frac{1}{2}\sum_{k}\Psi_{k}^{\dag}
[\mathcal{H}_{N}(\bk)+\mathcal{H}_{SC}(\bk)]\Psi_{k}$, where $\Psi_{k}$ denotes the Nambu basis,
$\mathcal{H}_{N}$ describes the normal state, and $H_{SC}$
describes the superconducting pairing. When $\mathcal{H}_{N}$ describes a
first-order topological insulator with $d>1$, one knows that
helical states will appear on the boundary and form $(d-1)$-dimensional
Dirac points at TRIMs
of the boundary Brillouin zone~\cite{hasan2010,qi2011}. If the chemical
potential is set to locate at the Dirac point,
then the low-energy Hamiltonian near the Dirac point for a given boundary
terminating at a given type of sublattice will take the standard form~\cite{shen2013topological}
\begin{eqnarray}
\mathcal{H}_{\scriptscriptstyle\boldsymbol{\Gamma}_{s}}(\bq)=\sum_{i=1}^{d-1}v_{i}q_{i}\gamma_{i},
\end{eqnarray}
where $\boldsymbol{\Gamma}_{s}$ denotes the TRIM at which the boundary
Dirac point is located, $\bq$ is the momentum measured from
$\boldsymbol{\Gamma}_{s}$, and the $\gamma_{i}$ matrices satisfy
the Clifford algebra, i.e., $\{\gamma_{i},\gamma_{j}\}=2\delta_{ij}$.
The effect from the superconducting pairing to the helical
states can be determined by projecting $\mathcal{H}_{SC}$ onto the  subspace spanned
by the orthogonal wave functions of helical boundary states. In general,
if one only considers the leading-order contribution,
what the superconducting pairing induces is a constant Dirac mass term to gap out
the Dirac point. Accordingly, the low-energy physics
on the boundary is described by a massive Dirac Hamiltonian of the form
\begin{eqnarray}
\tilde{\mathcal{H}}_{\scriptscriptstyle\boldsymbol{\Gamma}_{s}}(\bq)=\sum_{i=1}^{d-1}v_{i}q_{i}\gamma_{i}+m_{\scriptscriptstyle\boldsymbol{\Gamma}_{s}}\gamma_{d},
\end{eqnarray}
with $\{\gamma_{d},\gamma_{i=1,...,d-1}\}=0$.
Mathematically, the Dirac mass term is given by
\begin{eqnarray}
[m_{\scriptscriptstyle\boldsymbol{\Gamma}_{s}}\gamma_{d}]_{\alpha\beta}=\int dx_{d}\psi_{\alpha}^{\dag}(x_{d})\mathcal{H}_{\rm SC}(-i\partial_{x_{d}},\boldsymbol{\Gamma}_{s})
\psi_{\beta}(x_{d}),
\end{eqnarray}
where $\{\psi_{\alpha}(x_{d})\}$ denote the wave functions
for the helical states localized at the $x_{d}$-normal
boundary\cite{supplemental}. Focusing on the same boundary, if the location of the boundary
Dirac point changes from $\boldsymbol{\Gamma}_{s}$
to $\boldsymbol{\Gamma}_{s'}$ due to a change of the terminating
sublattice type, then the
boundary Hamiltonian will accordingly change to
\begin{eqnarray}
\tilde{\mathcal{H}}_{\scriptscriptstyle \boldsymbol{\Gamma}_{s'}}(\bq')=\sum_{i=1}^{d-1}v_{i}'q_{i}'\gamma_{i}+m_{\scriptscriptstyle\boldsymbol{\Gamma}_{s'}}\gamma_{d},
\end{eqnarray}
where $\bq'$ denotes the momentum measured from $\boldsymbol{\Gamma}_{s'}$.
While the value of Fermi velocity for the helical states on a given
boundary may also change, the sign
cannot change as each branch of the helical states must propagate in a fixed
direction. However,  the superconductivity-induced Dirac mass can change its magnitude as well as the sign if the
pairing has a momentum dependence, e.g., extended s-wave
pairing,  d-wave pairing etc. Without loss of generality, let us
now consider a nonuniform boundary consisting of two parts which respectively terminate at two
distinct types of sublattices. For the convenience of discussion, we dub 
the interface separating two distinct types of terminating sublattices as {\it sublattice domain wall}. 
Assuming that the sublattice domain walls only break the
translation symmetry of the given boundary in the $x_{d-1}$ direction,
the boundary Hamiltonian becomes
\begin{eqnarray}
\mathcal{H}(-i\partial_{x_{d-1}},\bq'_{\parallel})&=&-iv_{d-1}(x_{d-1})\gamma_{d-1}\partial_{x_{d-1}}
+m(x_{d-1})\gamma_{d}\nonumber\\
&&+\sum_{i=1}^{d-2}v_{i}q_{i}'\gamma_{i},
\end{eqnarray}
where $\bq'_{\parallel}=(q_{1}',...,q_{d-2}')$ denotes the momentum parallel to the sublattice domain walls.
Notably, if $m_{\scriptscriptstyle \boldsymbol{\Gamma}_{s}}$ and $m_{\scriptscriptstyle \boldsymbol{\Gamma}_{s'}}$ have opposite signs,
then the Dirac mass
$m(x_{d-1})$ will change sign across the sublattice domain walls. In other words,
the sublattice domain walls are domain walls of Dirac mass. As a result,  Majorana modes
with $d_{c}=2$ will emerge at the sublattice domain walls according to the Jackiw-Rebbi theory~\cite{jackiw1976b},
corresponding to the realization of an extrinsic time-reversal
invariant second-order TSC. As TRS
is conserved, the resulting Majorana modes will
be Majorana Kramers pairs (two MZMs related by TRS)
in two dimensions~\cite{Yan2018hosc,Wang2018hosc} and propagating helical Majorana modes
in three dimensions~\cite{Zhang2019hinge}.

The above general theory can be straightforwardly generalized to systems without TRS.
Without loss of generality, let us consider that the TRS
is broken by an external magnetic field. As Dirac mass induced by superconductivity and Zeeman field
will compete, if $|m_{\scriptscriptstyle\boldsymbol{\Gamma}_{s}}|\neq |m_{\scriptscriptstyle\boldsymbol{\Gamma}_{s'}}|$
and the absolute value of the Zeeman-field-induced
Dirac mass falls between $|m_{\scriptscriptstyle\boldsymbol{\Gamma}_{s}}|$ and $|m_{\scriptscriptstyle\boldsymbol{\Gamma}_{s'}}|$,
the Dirac mass of domain walls will become dominated by Zeeman field on one side and by superconductivity on the other side~\cite{supplemental}.
As a result, the  Majorana Kramers pairs and helical
Majorana modes will respectively change to single MZMs and chiral Majorana modes, with
their locations still bound at the sublattice domain walls~\cite{supplemental}.
With the established general theory in mind,
below we consider a concrete realization to
demonstrate the discussed physics.

{\it Kane-Mele model with spin-singlet pairing.---}
Since two-dimensional honeycomb lattices with just two types of sublattices
allow a simple illustration of the essential physics,
below we consider the representative Kane-Mele model to describe the 
topological insulator and further assume a proximity-induced spin-singlet 
pairing. The full Hamiltonian has the form
\begin{eqnarray}
H&=&t\sum_{\langle ij\rangle,\alpha}c_{i,\alpha}^{\dag}c_{j,\alpha}+i\lambda_{\rm so}\sum_{\langle\langle ij\rangle\rangle,\alpha,\beta}\nu_{ij}c_{i,\alpha}^{\dag}(s^{z})_{\alpha\beta}c_{j,\beta}\nonumber\\
&&-\mu\sum_{i,\alpha}c_{i,\alpha}^{\dag}c_{i,\alpha}+[\Delta_{0}\sum_{i}c_{i,\uparrow}^{\dag}c_{i,\downarrow}^{\dag}
+\sum_{\langle ij\rangle} \Delta_{1;ij}c_{i,\uparrow}^{\dag}c_{j,\downarrow}^{\dag}\nonumber\\
&&+\sum_{\langle\langle ij\rangle\rangle} \Delta_{2;ij} c_{i,\uparrow}^{\dag}c_{j,\downarrow}^{\dag}+h.c.],
\end{eqnarray}
where $\langle ij\rangle$ and $\langle\langle ij\rangle\rangle$ refer to nearest-neighbor
and next-nearest-neighbor sites.
The first line corresponds to the Kane-Mele model which realizes a
two-dimensional first-order topological insulator as long as the spin-orbit
coupling coefficient $\lambda_{\rm so}$ is nonzero~\cite{Kane2005a,Kane2005b}. $\mu$ is the chemical potential, $\Delta_{0}$,
$\Delta_{1;ij}$ and $\Delta_{2;ij}$ represent the on-site, nearest-neighbor and next-nearest-neighbor
pairings, respectively.  To have momentum dependence in
the pairing, at least one of $\Delta_{1;ij}$ and $\Delta_{2;ij}$ needs to be nonzero.
It is worth noting that according to the general theory,
there is no constraint on the pairing type (a demonstration of the physics
via d-wave pairing is  provided in the supplemental material~\cite{supplemental}).  Without loss of generality,
below we assume $\Delta_{1;ij}=\Delta_{1}$ and $\Delta_{2;ij}=\Delta_{2}$ for simplicity,
corresponding to an extended s-wave pairing which
preserves all crystalline symmetry of the normal-state Hamiltonian.

By a Fourier transformation to the momentum space and choosing
the basis to be $\Psi_{k}^{\dag}
=(\psi_{k}^{\dag},\psi_{-k})$ with $\psi_{k}^{\dag}=(c_{A,k,\uparrow}^{\dag},c_{B,k,\uparrow}^{\dag},
c_{A,k,\downarrow}^{\dag},c_{B,k,\downarrow}^{\dag})$, the BdG Hamiltonian reads
\begin{eqnarray}
\mathcal{H}(\bk)&=&t\sum_{i}[\cos(\bk\cdot\ba_{i})\tau_{z}s_{0}\sigma_{x}+\sin(\bk\cdot\ba_{i})\tau_{z}s_{0}\sigma_{y}]\nonumber\\
&&+2\lambda_{\rm so}\sum_{i}\sin(\bk\cdot\bb_{i})\tau_{0}s_{z}\sigma_{z}-\mu\tau_{z}s_{0}\sigma_{0}\nonumber\\
&&-\Delta_{1}\sum_{i}[\cos(\bk\cdot\ba_{i})\tau_{y}s_{y}\sigma_{x}+\sin(\bk\cdot\ba_{i})\tau_{y}s_{y}\sigma_{y}]\nonumber\\
&&-[\Delta_{0}+2\Delta_{2}\sum_{i}\cos(\bk\cdot\bb_{i})]\tau_{y}s_{y}\sigma_{0},\label{BdG}
\end{eqnarray}
where the Pauli matrices $\tau_{i}$, $s_{i}$ and $\sigma_{i}$ act on
the particle-hole, spin ($\uparrow,\downarrow$) and sublattice $(A,B)$
degrees of freedom, respectively. The sum runs over $i=1,2,3$, with
the nearest-neighbor vectors $\ba_{1}=a(0,1)$, $\ba_{2}=\frac{a}{2}(\sqrt{3},-1)$,
$\ba_{3}=\frac{a}{2}(-\sqrt{3},-1)$, and
$a$ being the lattice constant (below we set $a=1$ for notational
simplicity). The next-nearest-neighbor vectors
$\bb_{1}=\ba_{2}-\ba_{3}$, $\bb_{2}=\ba_{3}-\ba_{1}$ and $\bb_{3}=\ba_{1}-\ba_{2}$~\cite{Haldane1988}.
The Hamiltonian has TRS (the time-reversal operator $\mathcal{T}=i\tau_{0}s_{y}\sigma_{0}\mathcal{K}$
with $\mathcal{K}$ the complex conjugate operator), particle-hole
symmetry ($\mathcal{P}=\tau_{x}s_{0}\sigma_{0}\mathcal{K}$),
and inversion symmetry ($I=\tau_{0}s_{0}\sigma_{x}$). Because the
coexistence of TRS and inversion symmetry enforces
Kramers degeneracy to the bulk bands,   the first-order topology of the
BdG Hamiltonian will always be trivial for the
concerned spin-singlet pairing~\cite{Li2021bts,Majid2021surface,qi2010d}.  In previous works, it has been shown
that a topological insulator with square lattice in proximity
to an extended s-wave superconductor can realize a second-order
TSC with Majorana Kramers pairs localized
at the corners of a square sample~\cite{Yan2018hosc}. Notably, therein the topological criterion
requires either the hopping or the pairing to have crystalline anisotropy,
because otherwise domain walls of Dirac mass cannot form on the boundary
due to symmetry constraint. However, as we will show below, even though both the hopping and pairing
are considered to be isotropic in Eq.(\ref{BdG}), here domain walls
of Dirac mass can still
form on the boundary due to the sublattice degrees of freedom.

\begin{figure}[t]
\centering
\includegraphics[width=0.5\textwidth]{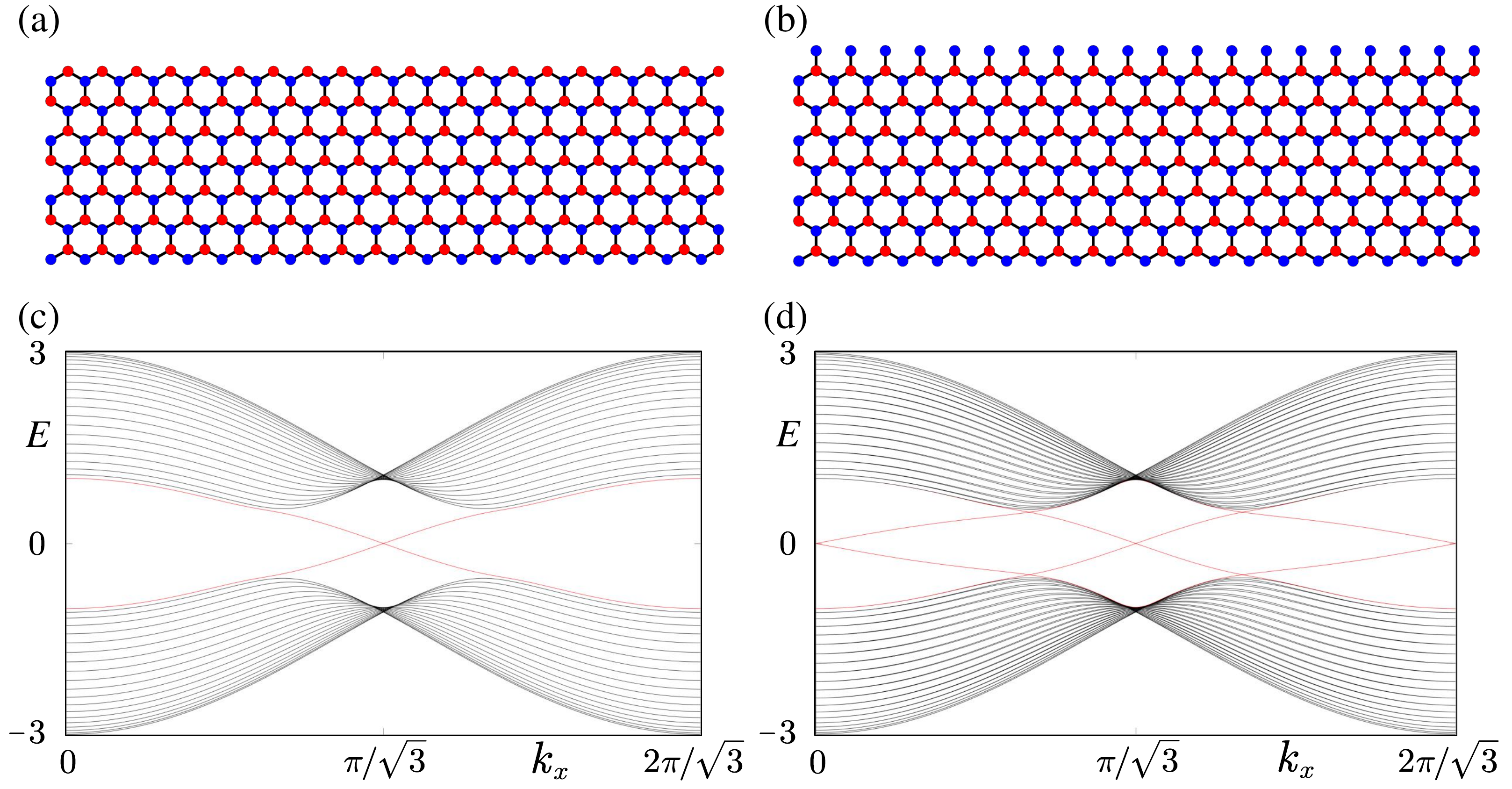}
\caption{(Color online) The sensitive dependence of
boundary Dirac points on the terminating sublattice type.
(a) The upper and lower zigzag edges of the lattice respectively
terminate at sublattice B (red dots) and A (blue dots). (b) The lower edge keeps to be the same
as in (a), but the upper edge changes to be a beard type, with the terminating sublattice type
changing from B to A.  (c) and (d) show the corresponding normal-state energy spectra
when the $y$-normal open boundaries  follow the structures shown in (a) and (b), respectively.
In (c)(d), periodic boundary condition is assumed in the $x$ direction,
and parameters are $t=1$, $\lambda_{so}=0.1$.}\label{DP}
\end{figure}

For the honeycomb lattice, there are two kinds of simple boundaries
whose outermost sublattices only contain one type, which are known
as zigzag and beard boundaries (see Figs.\ref{DP}(a)(b)). Let us first investigate the influence
of the change of terminating sublattice type on a given boundary to the helical
edge states of the normal state. To be specific,
we consider a cylindrical geometry with periodic boundary
condition in the $x$ direction and open boundary condition in the $y$ direction.
When the upper edge terminates at type-B sublattices and the lower edge
terminates at type-A sublattices (see Fig.\ref{DP}(a)),  one finds that the
boundary Dirac points for both upper and lower edges
are located at $k_{x}=\pi/\sqrt{3}$,  as shown in Fig.\ref{DP}(c).
By only changing the terminating sublattice type on the upper edge, one finds that one
boundary Dirac point is immediately shifted
from $k_{x}=\pi/\sqrt{3}$ to $k_{x}=0$, as shown in Figs.\ref{DP}(b)(d). Since nothing
changes in the bulk as well as on the lower edge, the shifted Dirac point
apparently corresponds to the upper edge, indicating the
sensitive sublattice-dependence of boundary Dirac points.

\begin{figure}[t]
\centering
\includegraphics[width=0.5\textwidth]{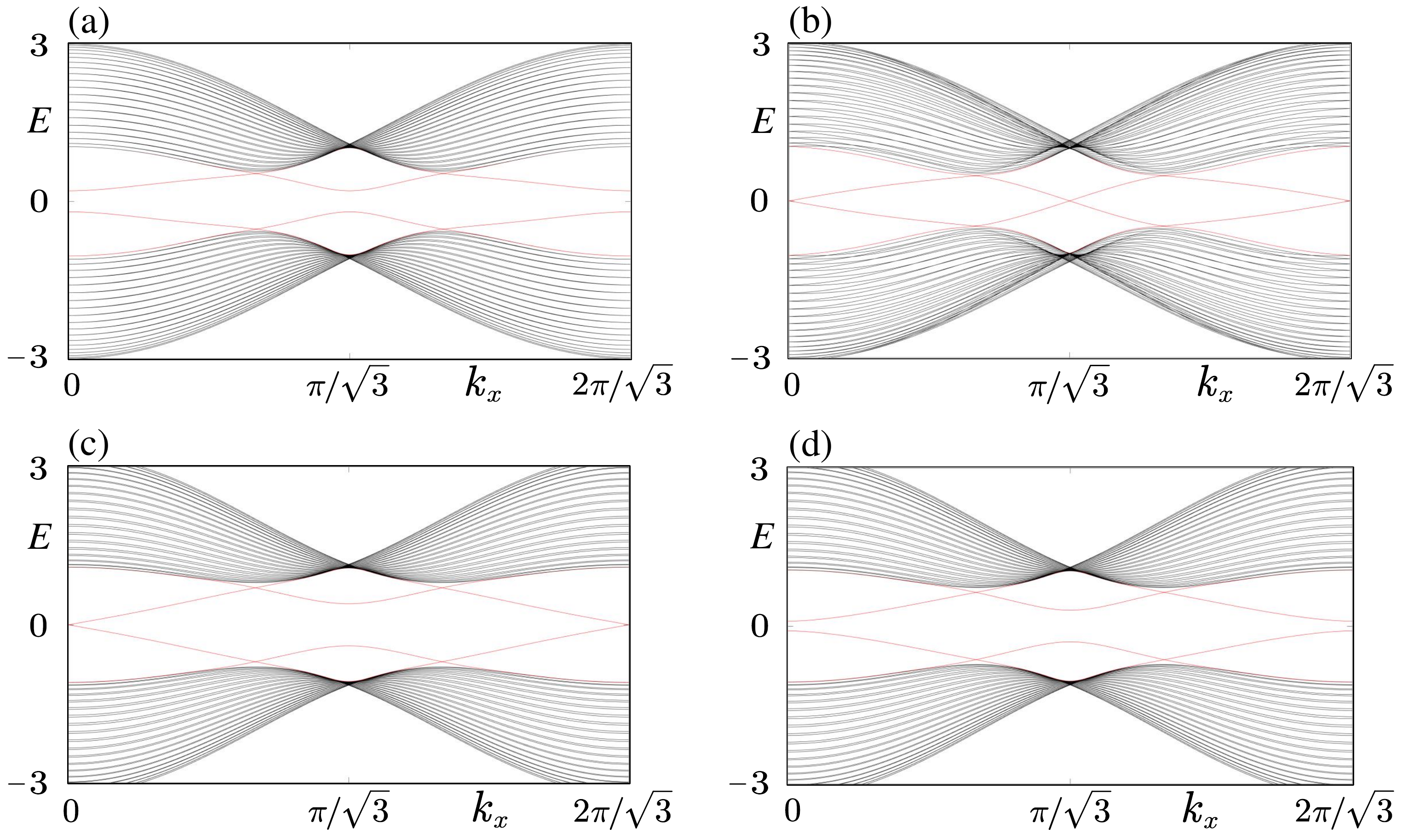}
\caption{(Color online) The energy spectrum of the BdG Hamiltonian for a cylindrical geometry with
open (periodic) boundary condition in the $y(x)$ direction. The upper (lower) edge in the $y$ direction is
 chosen to be the beard (zigzag) type. In (a)-(d), $t=1$, $\lambda_{so}=0.1$, $\mu=0$, and pairing
 amplitudes are: (a) $\Delta_{0}=0.2$, $\Delta_{1}=\Delta_{2}=0$;
(b) $\Delta_{1}=0.2$, $\Delta_{0}=\Delta_{2}=0$; (c) $\Delta_{2}=0.2$, $\Delta_{0}=\Delta_{1}=0$;
(d) $\Delta_{0}=0.1$, $\Delta_{1}=0$, $\Delta_{2}=0.2$.}\label{gap}
\end{figure}

Taking into account the superconductivity,
numerical results show that
the on-site pairing, nearest-neighbor pairing and
next-nearest-neighbor pairing have rather different effects to
the helical edge states, as shown in Fig.\ref{gap}. The on-site pairing,
as expected, will induce a Dirac mass to gap out the Dirac points, irrespective
of whether the edge is zigzag-type or beard-type, as shown in Fig.\ref{gap}(a). In sharp contrast,
Fig.\ref{gap}(b) shows that the boundary Dirac points are intact to the nearest-neighbor pairing.
Last, the next-nearest-neighbor pairing turns out to open a gap for the Dirac point
of the zigzag boundary but not for that of the beard boundary, as shown in Fig.\ref{gap}(c).
These results indicate when both $\Delta_{0}$
and $\Delta_{2}$ are finite, the gaps opened for the Dirac points at $k_{x}=0$ and
$k_{x}=\pi/\sqrt{3}$ can be different, as shown in Fig.\ref{gap}(d).

As the effect of the nearest-neighbor pairing to
the helical edge states is negligible,  below we set $\Delta_{1}=0$
for simplicity. To obtain the topological criterion
for the emergence of domain walls binding Majorana modes,
we follow the general theory and derive the low-energy
boundary Hamiltonians for both zigzag and beard  edges~\cite{supplemental}.
Focusing on the upper $y$-normal boundary and considering the case with $\mu=0$
for illustration, we find that
the boundary Hamiltonian for the beard-type edge (terminating at type-A sublattices) is
\begin{eqnarray}
\mathcal{H}_{\rm u,beard}(q_{x})=vq_{x}\tau_{0}s_{z}-\Delta_{0}\tau_{y}s_{y},
\end{eqnarray}
where $v=3\sqrt{3}\lambda_{so}$ and $q_{x}$ is measured from $k_{x}=0$,
and the boundary Hamiltonian for the zigzag-type edge (terminating at type-B sublattices)
is
\begin{eqnarray}
\mathcal{H}_{\rm u,zigzag}(q_{x}')=v'q_{x}'\tau_{0}s_{z}+(2\Delta_{2}-\Delta_{0})\tau_{y}s_{y},
\end{eqnarray}
where $v'\approx6\sqrt{3}\lambda_{so}$ if $\lambda_{so}/t\ll1$, and $q_{x}'$ is measured
from $k_{x}=\pi/\sqrt{3}$~\cite{supplemental}. It is easy to find that
the Dirac masses in the two Hamiltonians will take opposite signs
if $|\Delta_{2}|>|\Delta_{0}|/2>0$. This is the topological criterion
for sublattice domain walls to host Majorana Kramers pairs at $\mu=0$.
Due to the robustness of topology, this topological criterion will hold
as long as $\mu$ is lower than the critical value at which the boundary
energy gap gets closed~\cite{supplemental}.

\begin{figure}[t]
\centering
\includegraphics[scale=0.4]{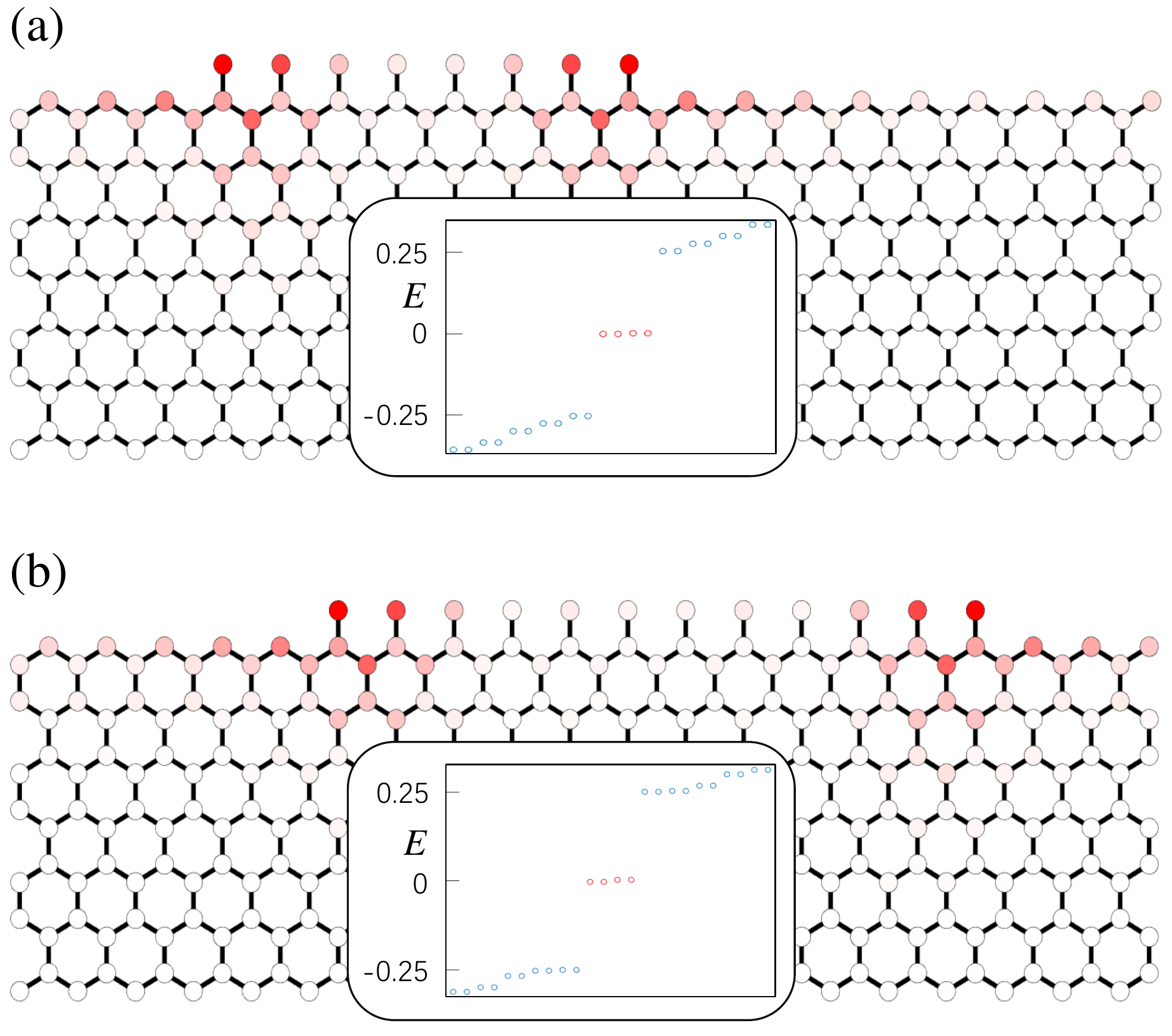}
\caption{(Color online) Majorana Kramers pairs bound at sublattice domain walls.
Parameters in (a) and (b) are $t=1$, $\lambda_{so}=0.1$, $\mu=0$, $\Delta_{0}=\Delta_{2}=0.3$,
$\Delta_{1}=0$. With periodic
boundary condition in the $x$ direction except for the uppermost beard-type part,
the two insets in (a) and (b) show
the corresponding energy spectra. The four dots highlighted by red indicate
the existence of two Majorana Kramers pairs.  The shade of the red color
on the lattice sites reflect the weight of the probability density of Majorana Kramers pairs. }\label{domain}
\end{figure}

To validate the established topological criterion, we
still consider a cylindrical geometry with periodic boundary condition
in the $x$ direction and just let the upper edge be nonuniform,
with one part terminating at B-type sublattices (zigzag) and the other part
terminating at A-type sublattices (beard). Accordingly, there are two sublattice domain walls
on the upper edge, while the lower edge keeps uniform.
As shown in Fig.\ref{domain}, when the topological criterion is fulfilled, a diagonalization
of the Hamiltonian shows the existence of four MZMs, corresponding to two
Majorana Kramers pairs. As expected, the wave functions of Majorana Kramers
pairs are strongly localized around the sublattice domain walls. In addition, by
a comparison of Figs.\ref{domain}(a) and (b), it is readily seen that
the positions of Majorana Kramers pairs directly follow
the change of the positions of sublattice domain walls,
indicating that the positions of Majorana Kramers pairs can be tuned site-by-site
by a precise control of the terminating sublattices.
Remarkably, even when the positions of sublattice domain walls are fixed,
we find that the same goal can also be achieved by electrically tuning the
local boundary potential~\cite{supplemental}.

{\it Discussion and conclusion.---} While our theory is exemplified
in terms of the two-dimensional honeycomb lattice, its generality admits a wide
application as sublattice degrees of freedom are rather common
in materials, e.g., the class of materials with kagome or Lieb lattice consist
of three types of sublattices~\cite{Guo2009,Week2010}. As a new scheme for the implementation
of extrinsic second-order TSCs and Majorana modes, one remarkable merit is that the sensitive sublattice-dependence
allows the positions of Majorana modes to be manipulated at an atomically precise level. Therefore, our work
opens a promising avenue for achieving the manipulation and braiding of Majorana modes.

{\it Acknowledgements.---} We thank  Zhi Wang for helpful discussions.
This work is supported by the National Natural Science Foundation of China (Grant No.11904417
and 12174455) and the Natural Science Foundation of Guangdong Province
(Grant No. 2021B1515020026).

\bibliography{dirac}

\begin{widetext}
\clearpage
\begin{center}
\textbf{\large Supplemental Material ``Sublattice-sensitive Majorana Modes''}\\
\vspace{4mm}
{Di Zhu$^{1,*}$, Bo-Xuan Li$^{2,3,*}$, Zhongbo Yan$^{1,\dagger}$}\\
\vspace{2mm}
{\em $^1$School of Physics, Sun Yat-Sen University, Guangzhou, 510275, China}\\
{\em $^2$Beijing National Laboratory for Condensed Matter Physics and Institute of Physics,
Chinese Academy of Sciences, Beijing 100190, China}\\
{\em $^3$University of Chinese Academy of Sciences, Beijing 100049, China}
\end{center}

\setcounter{equation}{0}
\setcounter{figure}{0}
\setcounter{table}{0}
\makeatletter
\renewcommand{\theequation}{S\arabic{equation}}
\renewcommand{\thefigure}{S\arabic{figure}}
\renewcommand{\bibnumfmt}[1]{[S#1]}

This supplemental material contains five sections, including: (I)
the derivation of  low-energy boundary Hamiltonians for beard and zigzag edges;
(II) two-dimensional honeycomb-lattice topological insulators in proximity to d-wave superconductors;
(III) the impact of finite chemical potential on the topological criterion;
(IV) sublattice-sensitive Majorana modes in time-reversal symmetry broken systems;
(V) manipulating the positions of Majorana zero modes by electrically controlling
the local boundary potential.


\section{I. The derivation of low-energy boundary Hamiltonians for beard and zigzag edges}

To derive the low-energy boundary Hamiltonian, we write down the bulk
Bogoliubov-de Gennes (BdG) Hamiltonian in an explicit form, which reads
\begin{eqnarray}
\mathcal{H}_{\rm BdG}(\bk)&=&t(2\cos\frac{\sqrt{3}k_{x}}{2}\cos\frac{k_{y}}{2}+\cos k_{y})\tau_{z}s_{0}\sigma_{x}
-t(2\cos\frac{\sqrt{3}k_{x}}{2}\sin\frac{k_{y}}{2}-\sin k_{y})\tau_{z}s_{0}\sigma_{y}\nonumber\\
&&+2\lambda_{so}(\sin\sqrt{3}k_{x}-2\sin\frac{\sqrt{3}k_{x}}{2}\cos\frac{3k_{y}}{2})\tau_{0}s_{z}\sigma_{z}
-\mu\tau_{z}s_{0}\sigma_{0}\nonumber\\
&&-\Delta_{1}(2\cos\frac{\sqrt{3}k_{x}}{2}\cos\frac{k_{y}}{2}+\cos k_{y})\tau_{y}s_{y}\sigma_{x}
+\Delta_{1}(2\cos\frac{\sqrt{3}k_{x}}{2}\sin\frac{k_{y}}{2}-\sin k_{y})\tau_{y}s_{y}\sigma_{y}\nonumber\\
&&-[\Delta_{0}+2\Delta_{2}(\cos\sqrt{3}k_{x}+2\cos\frac{\sqrt{3}k_{x}}{2}\cos\frac{3k_{y}}{2})]\tau_{y}s_{y}\sigma_{0}.\label{BdGH}
\end{eqnarray}
For notational simplicity, we have set the lattice constant to unity.
For the convenience of discussion, throughout this work, $t$ and $\lambda_{so}$
will be assumed to be positive.
As we have shown numerically that the nearest-neighbor pairing has a negligible effect
to the helical edge states, below we also set $\Delta_{1}=0$ for simplicity.

According to numerical results, we know that for a cylindrical geometry with periodic
boundary condition in the $x$ direction and open boundary condition in the $y$ direction,
the boundary Dirac point, which corresponds to the crossing point of the energy spectrum
of the normal-state helical edge states,  is located at $k_{x}=0$ ($k_{x}=\pi/\sqrt{3}$) for a beard (zigzag) edge.
Below let us focus on the upper $y$-normal
boundary and derive the corresponding low-energy boundary Hamiltonians for both beard and zigzag
edges.

\subsection{A. Low-energy boundary Hamiltonian for the beard edge}

When the upper boundary is a beard edge, the terminating sublattices are type A.
As numerical calculations reveal that the boundary Dirac point for such
an edge will appear at $k_{x}=0$, in order to derive the low-energy boundary
Hamiltonian, we perform an expansion of the bulk Hamiltonian around $k_{x}=0$ up to
the linear order in momentum. Accordingly, the Hamiltonian becomes
\begin{eqnarray}
\mathcal{H}_{\rm BdG}(q_{x},k_{y})&=&t(2\cos\frac{k_{y}}{2}+\cos k_{y})\tau_{z}s_{0}\sigma_{x}
-t(2\sin\frac{k_{y}}{2}-\sin k_{y})\tau_{z}s_{0}\sigma_{y}\nonumber\\
&&+2\sqrt{3}\lambda_{so}q_{x}(1-\cos\frac{3k_{y}}{2})\tau_{0}s_{z}\sigma_{z}
-\mu\tau_{z}s_{0}\sigma_{0}\nonumber\\
&&-[\Delta_{0}+2\Delta_{2}(1+2\cos\frac{3k_{y}}{2})]\tau_{y}s_{y}\sigma_{0},
\end{eqnarray}
where $q_{x}$ denotes a small momentum measured from $k_{x}=0$.
Next, we decompose the Hamiltonian into two parts, $\mathcal{H}_{\rm BdG}=\mathcal{H}_{1}+\mathcal{H}_{2}$,
with
\begin{eqnarray}
\mathcal{H}_{1}(q_{x},k_{y})&=&t(2\cos\frac{k_{y}}{2}+\cos k_{y})\tau_{z}s_{0}\sigma_{x}
-t(2\sin\frac{k_{y}}{2}-\sin k_{y})\tau_{z}s_{0}\sigma_{y}, \nonumber\\
\mathcal{H}_{2}(q_{x},k_{y})&=&2\sqrt{3}\lambda_{so}q_{x}(1-\cos\frac{3k_{y}}{2})\tau_{0}s_{z}\sigma_{z}
-\mu\tau_{z}s_{0}\sigma_{0}-[\Delta_{0}+2\Delta_{2}(1+2\cos\frac{3k_{y}}{2})]\tau_{y}s_{y}\sigma_{0}.
\end{eqnarray}
For real materials, $\lambda_{so}\ll t$ and $\Delta_{0,2}\ll t$ are naturally satisfied.
As we are interested in the regime where $q_{x}$ is small, the whole $\mathcal{H}_{2}$
can be treated as a perturbation if the chemical potential is also
assumed to be close to the neutrality condition.

In the following, let us consider a half-infinity sample with the boundary corresponding
to the upper beard edge. In the basis $\Psi_{q_{x}}=(c_{1,A,q_{x}},c_{1,B,q_{x}},c_{2,A,q_{x}},c_{2,B,q_{x}},
...,c_{n,A,q_{x}},c_{n,B,q_{x}},...)^{T}$ with $c_{n,A(B),q_{x}}=
(c_{n,A(B),q_{x},\uparrow},c_{n,A(B),q_{x},\downarrow},c_{n,A(B),-q_{x},\uparrow}^{\dag},c_{n,A(B),-q_{x},\downarrow}^{\dag})$,
the Hamiltonian in the matrix form reads
\begin{eqnarray}
\mathcal{H}_{1}=\left(
        \begin{array}{ccccccc}
          0 & t\tau_{z}s_{0} & 0 & 0 & 0 & 0 & \cdots \\
          t\tau_{z}s_{0} & 0 & 2t\tau_{z}s_{0} & 0 & 0 & 0 & \cdots \\
          0 & 2t\tau_{z}s_{0} & 0 & t\tau_{z}s_{0} & 0 & 0 & \cdots \\
          0 & 0 & t\tau_{z}s_{0} & 0 & 2t\tau_{z}s_{0} & 0 & \cdots \\
          0 & 0 & 0 & 2t\tau_{z}s_{0} & 0  & t\tau_{z}s_{0} & \cdots \\
          0 & 0 & 0 & 0 & t\tau_{z}s_{0} & 0 & \cdots \\
          \vdots & \vdots & \vdots & \vdots & \vdots & \vdots & \ddots \\
        \end{array}
      \right).
\end{eqnarray}
To see that this Hamiltonian has solutions for zero-energy bound states,
we solve the eigenvalue equation $\mathcal{H}_{1}|\Psi_{\alpha}\rangle=0$. Concretely,
as $\tau_{z}$ and $s_{z}$ both commute with $\mathcal{H}_{1}$, the eigenvector
$|\Psi_{\alpha}\rangle$ can be assigned with the form
\begin{eqnarray}
|\Psi_{\tau s}\rangle=|\tau_{z}=\tau\rangle\otimes|s_{z}=s\rangle\otimes(\psi_{1A},\psi_{1B},\psi_{2A},\psi_{2B},...,
\psi_{nA},\psi_{nB},...)^{T},
\end{eqnarray}
where $\tau=\pm1$ and $s=\pm1$ correspond to the two possible eigenvalues of $\tau_{z}$
and $s_{z}$, respectively. Taking the expression of $|\Psi_{\tau s}\rangle$ back into
the eigenvalue equation $\mathcal{H}_{1}|\Psi_{\tau s}\rangle=0$, one gets a series of
equations with periodic structures, which read
\begin{eqnarray}
&&t_{\tau}\psi_{1B}=0,\nonumber\\
&&t_{\tau}\psi_{1A}+2t_{\tau}\psi_{2A}=0,\nonumber\\
&&2t_{\tau}\psi_{1B}+t_{\tau}\psi_{2B}=0,\nonumber\\
&&...\nonumber\\
&&t_{\tau}\psi_{nA}+2t_{\tau}\psi_{(n+1)A}=0,\nonumber\\
&&2t_{\tau}\psi_{nB}+t_{\tau}\psi_{(n+1)B}=0,\nonumber\\
&&...,
\end{eqnarray}
where $t_{\tau}\equiv t\tau$. According to the periodic structures, one can easily find
\begin{eqnarray}
\psi_{(n+1)A}=-\frac{1}{2}\psi_{nA}, \quad \psi_{nB}=0.
\end{eqnarray}
Therefore, the eigenvectors take the form
\begin{eqnarray}
|\Psi_{\tau s}\rangle=|\tau_{z}=\tau\rangle\otimes|s_{z}=s\rangle\otimes\mathcal{N}(1,0,-\frac{1}{2},0,...,
(-\frac{1}{2})^{(n-1)},0,...)^{T},
\end{eqnarray}
where the normalization constant $\mathcal{N}$ is determined by
the normalization condition
$\langle \Psi_{\tau s} |\Psi_{\tau s}\rangle=1$. Simple algebra calculations give
\begin{eqnarray}
\mathcal{N}^{2}\sum_{n=0}^{\infty}\frac{1}{2^{2n}}=\mathcal{N}^{2}\frac{1}{1-\frac{1}{4}}=\frac{4}{3}\mathcal{N}^{2}=1,
\end{eqnarray}
indicating  $\mathcal{N}=\frac{\sqrt{3}}{2}$. As $\psi_{nA}$ decays in a power law with the increase of $n$,
the existence of four such eigenvectors indicates the existence of four zero-energy bound states. It is worth noting
that the topological insulator has one pair of helical states on a given edge, but the introduce of
particle-hole degrees of freedom doubles the number of helical states.
Next, we project
$\mathcal{H}_{2}$ onto the basis spanned by the four zero-energy eigenvectors. To proceed, we write down the
matrix form for each term in $\mathcal{H}_{2}$.

Let us first focus on the term $\mathcal{H}_{2,1}=2\sqrt{3}\lambda_{so}q_{x}(1-\cos\frac{3k_{y}}{2})\tau_{0}s_{z}\sigma_{z}$.
In the basis $\Psi_{q_{x}}=(c_{1,A,q_{x}},c_{1,B,q_{x}},c_{2,A,q_{x}},c_{2,B,q_{x}},
...,c_{n,A,q_{x}},c_{n,B,q_{x}},...)^{T}$, its matrix form is
\begin{eqnarray}
\mathcal{H}_{2,1}=\tau_{0}\otimes s_{z}\otimes\left(
        \begin{array}{ccccccc}
          2\sqrt{3}\lambda_{so}q_{x} & 0 & -\sqrt{3}\lambda_{so}q_{x} & 0 & 0 & 0 & \cdots \\
          0 & -2\sqrt{3}\lambda_{so}q_{x} & 0 & \sqrt{3}\lambda_{so}q_{x} & 0 & 0 & \cdots \\
          -\sqrt{3}\lambda_{so}q_{x} & 0 & 2\sqrt{3}\lambda_{so}q_{x} & 0 & -\sqrt{3}\lambda_{so}q_{x} & 0 & \cdots \\
          0 & \sqrt{3}\lambda_{so}q_{x} & 0 & -2\sqrt{3}\lambda_{so}q_{x} & 0 & \sqrt{3}\lambda_{so}q_{x} & \cdots \\
          0 & 0 & -\sqrt{3}\lambda_{so}q_{x} & 0 & 2\sqrt{3}\lambda_{so}q_{x}  & 0 & \cdots \\
          0 & 0 & 0 & \sqrt{3}\lambda_{so}q_{x} & 0 & -2\sqrt{3}\lambda_{so}q_{x} & \cdots \\
          \vdots & \vdots & \vdots & \vdots & \vdots & \vdots & \ddots \\
        \end{array}
      \right).
\end{eqnarray}
Then the contribution from $\mathcal{H}_{2,1}$ to the boundary Hamiltonian is
\begin{eqnarray}
(\mathcal{H}_{beard,1})_{\tau s,\tau' s'}&=&\langle \Psi_{\tau s}|\mathcal{H}_{2,1}|\Psi_{\tau' s'}\rangle\nonumber\\
&=&2\sqrt{3}\lambda_{so,s}q_{x}\delta_{\tau\tau'}\delta_{ss'}
-\sqrt{3}\lambda_{so,s}q_{x}\delta_{\tau\tau'}\delta_{ss'}\mathcal{N}^{2}\sum_{n=1}^{+\infty}2\psi_{n A}\psi_{(n+1)A}\nonumber\\
&=&2\sqrt{3}\lambda_{so,s}q_{x}\delta_{\tau\tau'}\delta_{ss'}+\sqrt{3}\lambda_{so,s}q_{x}\delta_{\tau\tau'}\delta_{ss'}\mathcal{N}^{2}
\sum_{n=1}^{+\infty}\psi_{n A}\psi_{nA}\nonumber\\
&=&2\sqrt{3}\lambda_{so,s}q_{x}\delta_{\tau\tau'}\delta_{ss'}+\sqrt{3}\lambda_{so,s}q_{x}\delta_{\tau\tau'}\delta_{ss'}\nonumber\\
&=&3\sqrt{3}\lambda_{so,s}q_{x}\delta_{\tau\tau'}\delta_{ss'},
\end{eqnarray}
where $\lambda_{so,s}\equiv\lambda_{so}s$. In the derivation above, a few facts have been used, including: (1) $\psi_{nA}$ is real; (2)
$\psi_{(n+1)A}=-\frac{1}{2}\psi_{nA}$; (3)  $\mathcal{N}^{2}\sum_{n=1}^{+\infty}\psi_{nA}^{2}=1$.
Choosing the basis spanning the subspace for boundary Hamiltonian
to be $(|\Psi_{11}\rangle,|\Psi_{1-1}\rangle,|\Psi_{-11}\rangle,|\Psi_{-1-1}\rangle)^{T}$,
$\mathcal{H}_{beard,1}$ can be expressed in terms of the Pauli matrices as
\begin{eqnarray}
\mathcal{H}_{beard,1}=3\sqrt{3}\lambda_{so}q_{x}\tau_{0}s_{z}.
\end{eqnarray}

For the second term $\mathcal{H}_{2,2}=-\mu\tau_{z}s_{0}\sigma_{0}$, as it is diagonal
in the basis $\Psi_{q_{x}}$, one can easily find that its contribution to
the boundary Hamiltonian is just
\begin{eqnarray}
\mathcal{H}_{beard,2}=-\mu\tau_{z}s_{0}.
\end{eqnarray}
Now let us analyze the contribution from the pairing term. In the basis $\Psi_{q_{x}}$,
the matrix form of the pairing term is
\begin{eqnarray}
\mathcal{H}_{2,3}=-\left(
          \begin{array}{ccccc}
            (\Delta_{0}+2\Delta_{2})\tau_{y}s_{y} & 0 & 2\Delta_{2}\tau_{y}s_{y} & 0 & \cdots \\
            0 & (\Delta_{0}+2\Delta_{2})\tau_{y}s_{y} & 0 & 2\Delta_{2}\tau_{y}s_{y} & \cdots \\
            2\Delta_{2}\tau_{y}s_{y} & 0 & (\Delta_{0}+2\Delta_{2})\tau_{y}s_{y} & 0 & \cdots \\
            0 & 2\Delta_{2}\tau_{y}s_{y} & 0 & (\Delta_{0}+2\Delta_{2})\tau_{y}s_{y} & \cdots \\
            \vdots & \vdots & \vdots & \vdots & \ddots \\
          \end{array}
        \right).
\end{eqnarray}
Similarly, its contribution to the boundary Hamiltonian is
\begin{eqnarray}
(\mathcal{H}_{beard,3})_{\tau s,\tau' s'}&=&\langle \Psi_{\tau s}|\mathcal{H}_{2,3}|\Psi_{\tau' s'}\rangle\nonumber\\
&=&-(\Delta_{0}+2\Delta_{2})(\tau_{y})_{\tau\tau'}(s_{y})_{ss'}
-2\Delta_{2}(\tau_{y})_{\tau\tau'}(s_{y})_{ss'}\mathcal{N}^{2}\sum_{n=1}^{+\infty}2\psi_{n A}\psi_{(n+1)A}\nonumber\\
&=& -(\Delta_{0}+2\Delta_{2})(\tau_{y})_{\tau\tau'}(s_{y})_{ss'}+2\Delta_{2}(\tau_{y})_{\tau\tau'}(s_{y})_{ss'}\nonumber\\
&=&-\Delta_{0}(\tau_{y})_{\tau\tau'}(s_{y})_{ss'}.
\end{eqnarray}
One finds  that the contribution from the next-nearest-neighbor pairing will vanish for
the beard edge, in agreement with the numerical results shown in Fig.2 of the main text.
In terms of the Pauli matrices,  its form is just
\begin{eqnarray}
\mathcal{H}_{beard,3}=-\Delta_{0}\tau_{y}s_{y}.
\end{eqnarray}
Taking all contributions together, we reach the final expression of the boundary Hamiltonian
for the upper beard edge, which reads
\begin{eqnarray}
\mathcal{H}_{beard}=\mathcal{H}_{beard,1}+\mathcal{H}_{beard,2}+\mathcal{H}_{beard,3}
=vq_{x}\tau_{0}s_{z}-\mu\tau_{z}s_{0}-\Delta_{0}\tau_{y}s_{y},\label{beardH}
\end{eqnarray}
where $v=3\sqrt{3}\lambda_{so}$. In the limit $\mu=0$, the boundary Hamiltonian reduces to the form of Eq.(8) in the main text.
We find that the boundary energy gap at $k_{x}=0$ predicted by the low-energy boundary Hamiltonian
agree perfectly with the numerical results when only the on-site pairing or the next-nearest-neighbor pairing
is present. When both the on-site and the next-nearest-neighbor pairings are finite,
the boundary energy gap is found to be a little smaller than the predicted value $E_{g}=2|\Delta_{0}|$, but the agreement is
still very good at the neighborhood of the boundary Dirac point.

\subsection{B. Low-energy boundary Hamiltonian for the zigzag edge}

When the upper edge changes to terminate at type-B sublattices, so a zigzag edge,
numerical results show that the boundary Dirac point is shifted to $k_{x}=\pi/\sqrt{3}$.
In order to analytically derive the corresponding boundary Hamiltonian, we similarly
perform an expansion around $k_{x}=\pi/\sqrt{3}$ and keep the momentum
up to the linear order. Accordingly, the Hamiltonian becomes
\begin{eqnarray}
\mathcal{H}_{\rm BdG}(q_{x}',k_{y})&=&t(-\sqrt{3}q_{x}'\cos\frac{k_{y}}{2}+\cos k_{y})\tau_{z}s_{0}\sigma_{x}
+t(\sqrt{3}q_{x}'\sin\frac{k_{y}}{2}+\sin k_{y})\tau_{z}s_{0}\sigma_{y}\nonumber\\
&&+2\lambda_{so}(-\sqrt{3}q_{x}'-2\cos\frac{3k_{y}}{2})\tau_{0}s_{z}\sigma_{z}
-\mu\tau_{z}s_{0}\sigma_{0}\nonumber\\
&&-[\Delta_{0}+2\Delta_{2}(-1-\sqrt{3}q_{x}'\cos\frac{3k_{y}}{2})]\tau_{y}s_{y}\sigma_{0},
\end{eqnarray}
where $q_{x}'$ denotes a small momentum measured from $k_{x}=\pi/\sqrt{3}$.
Similar to the previous case, we decompose the Hamiltonian into two parts, $\mathcal{H}=\mathcal{H}_{1}+\mathcal{H}_{2}$, with
\begin{eqnarray}
\mathcal{H}_{1}(q_{x}',k_{y})&=&t\cos k_{y}\tau_{z}s_{0}\sigma_{x}+t\sin k_{y}\tau_{z}s_{0}\sigma_{y}-4\lambda_{so}\cos\frac{3k_{y}}{2}\tau_{0}s_{z}\sigma_{z},\nonumber\\
\mathcal{H}_{2}(q_{x}',k_{y})&=&-\sqrt{3}tq_{x}'\cos\frac{k_{y}}{2}\tau_{z}s_{0}\sigma_{x}
+\sqrt{3}tq_{x}'\sin\frac{k_{y}}{2}\tau_{z}s_{0}\sigma_{y}\nonumber\\
&&-2\sqrt{3}\lambda_{so}q_{x}'\tau_{0}s_{z}\sigma_{z}
-\mu\tau_{z}s_{0}\sigma_{0}\nonumber\\
&&-[\Delta_{0}+2\Delta_{2}(-1-\sqrt{3}q_{x}'\cos\frac{3k_{y}}{2})]\tau_{y}s_{y}\sigma_{0}.
\end{eqnarray}
As we are interested in the small $q_{x}'$ regime, it is also justified to treat the whole $\mathcal{H}_{2}$ as a perturbation.

When the upper edge becomes a zigzag one, the terminating sublattices become type B, so  the corresponding basis
for a half-infinity system becomes $\Psi_{q_{x}'}=(c_{1,B,q_{x}'},c_{2,A,q_{x}'},c_{2,B,q_{x}'},c_{3,A,q_{x}'},c_{3,B,q_{x}'},
...,c_{n,A,q_{x}'},c_{n,B,q_{x}'},...)^{T}$.
Then $\mathcal{H}_{1}$ in matrix form reads
\begin{eqnarray}
\mathcal{H}_{1}=\left(
        \begin{array}{ccccccc}
          0 & 0 & 2\lambda_{so}\tau_{0}s_{z} & 0 & 0 & 0 & \cdots \\
          0 & 0 & t\tau_{z}s_{0} & -2\lambda_{so}\tau_{0}s_{z} & 0 & 0 & \cdots \\
          2\lambda_{so}\tau_{0}s_{z} & t\tau_{z}s_{0} & 0 & 0 & 2\lambda_{so}\tau_{0}s_{z} & 0 & \cdots \\
          0 & -2\lambda_{so}\tau_{0}s_{z} & 0 & 0 & t\tau_{z}s_{0} & -2\lambda_{so}\tau_{0}s_{z} & \cdots \\
          0 & 0 & 2\lambda_{so}\tau_{0}s_{z} & t\tau_{z}s_{0} & 0  & 0 & \cdots \\
          0 & 0 & 0 & -2\lambda_{so}\tau_{0}s_{z} & 0 & 0 & \cdots \\
          \vdots & \vdots & \vdots & \vdots & \vdots & \vdots & \ddots \\
        \end{array}
      \right),
\end{eqnarray}
As $\tau_{z}$ and $s_{z}$ also commute with $\mathcal{H}_{1}$, the zero-energy eigenvectors of $\mathcal{H}_{1}$ can also be assigned the form
\begin{eqnarray}
|\Psi_{\tau s}\rangle=|\tau_{z}=\tau\rangle\otimes|s_{z}=s\rangle\otimes(\psi_{1B},\psi_{2A},\psi_{2B},\psi_{3A},\psi_{3B},...,
\psi_{nA},\psi_{nB},...)^{T}.
\end{eqnarray}
Accordingly, the eigenvalue equation $\mathcal{H}_{1}|\Psi_{\tau s}\rangle=0$ leads to the following
equations  with periodic structures,
\begin{eqnarray}
&&2\lambda_{so,s}\psi_{2B}=0,\nonumber\\
&&t_{\tau }\psi_{2B}-2\lambda_{so, s}\psi_{3A}=0,\nonumber\\
&&2\lambda_{so,s}\psi_{1B}+t_{\tau }\psi_{2A}+2\lambda_{so, s}\psi_{3B}=0,\nonumber\\
&&-2\lambda_{so, s}\psi_{2A}+t_{\tau }\psi_{3B}-2\lambda_{so, s}\psi_{4A}=0,\nonumber\\
&&...\nonumber\\
&&2\lambda_{so,s}\psi_{(n-1)B}+t_{\tau }\psi_{nA}+2\lambda_{so,s}\psi_{(n+1)B}=0,\nonumber\\
&&-2\lambda_{so, s}\psi_{nA}+t_{\tau}\psi_{(n+1)B}-2\lambda_{so,s}\psi_{(n+2)A}=0,\nonumber\\
&&....
\end{eqnarray}
It is readily found that the components of eigenvectors have $\psi_{(2n)B}=\psi_{(2n+1)A}=0$. Therefore, we only need to focus on the following equations,
\begin{eqnarray}
&&2\lambda_{so,s}\psi_{(2n-1)B}+t_{\tau}\psi_{(2n)A}+2\lambda_{so,s}\psi_{(2n+1)B}=0,\nonumber\\
&&-2\lambda_{so,s}\psi_{(2n)A}+t_{\tau}\psi_{(2n+1)B}-2\lambda_{so,s}\psi_{2(n+2)A}=0.
\end{eqnarray}
Consider the  trial function
\begin{eqnarray}
\left(
  \begin{array}{c}
    \psi_{(2n+1)B} \\
    \psi_{(2n+2)A} \\
  \end{array}
\right)
=\xi^{n}\left(
  \begin{array}{c}
    \psi_{1B} \\
    \psi_{2A} \\
  \end{array}
\right),
\end{eqnarray}
where $|\xi|<1$ is required so that the wave function decays in real space and corresponds to a
bound state. Accordingly, one finds that the series of equations reduce to two algebra equations, which read
\begin{eqnarray}
&&2\lambda_{so,s}\psi_{1B}+t_{\tau}\psi_{2A}+2\lambda_{so,s}\xi\psi_{1B}=0,\nonumber\\
&&-2\lambda_{so,s}\psi_{2A}+t_{\tau}\xi\psi_{1B}-2\lambda_{so,s}\xi\psi_{2A}=0.
\end{eqnarray}
By simple algebra, one finds
\begin{eqnarray}
&&\psi_{1B}=-\frac{t_{\tau}}{2\lambda_{so,s}(1+\xi)}\psi_{2A}, \nonumber\\
&&\frac{2\lambda_{so,s}(1+\xi)}{t_{\tau}\xi}=-\frac{t_{\tau}}{2\lambda_{so,s}(1+\xi)}.\label{eigenvector}
\end{eqnarray}
There are two solutions for $\xi$,
\begin{eqnarray}
\xi_{\pm}=\frac{-(t_{\tau}^{2}+8\lambda_{so,s}^{2})\pm\sqrt{t_{\tau}^{2}(t_{\tau}^{2}+16\lambda_{so,s}^{2})}}{8\lambda_{so,s}^{2}},\label{zlength}
\end{eqnarray}
however, only $\xi_{+}$ leads to decaying wave functions, so bound states. Taking $\xi_{+}$ back into Eq.(\ref{eigenvector}),
one finds
\begin{eqnarray}
\psi_{1B}=\frac{4t_{\tau}\lambda_{so,s}^{2}}{\lambda_{so,s}\left(t_{\tau}^{2}-\sqrt{t_{\tau}^{2}(t_{\tau}^{2}+16\lambda_{so, s}^{2})}\right)}\psi_{2A}\equiv\eta_{\tau s}\psi_{2A}=-\tau s|\eta_{\tau s}|\psi_{2A}.\label{ratio}
\end{eqnarray}
As $\tau$ and $s$ have four possible combinations,  there are also four eigenvectors corresponding to four zero-energy bound states. The eigenvectors
can also be expressed as
\begin{eqnarray}
|\Psi_{\tau s}\rangle=|\tau_{z}=\tau\rangle\otimes|s_{z}=s\rangle\otimes\mathcal{N}(\eta_{\tau s},1,0,0,\xi_{+}\eta_{\tau s},\xi_{+},0,0,\xi_{+}^{2}\eta_{\tau s},\xi_{+}^{2},...)^{T}.
\end{eqnarray}
The normalization condition $\langle\Psi_{\tau s}|\Psi_{\tau s}\rangle=1$ gives
\begin{eqnarray}
\mathcal{N}^{2}(1+\eta_{\tau s}^{2})\sum_{n=0}^{\infty}\xi_{+}^{2n}=\mathcal{N}^{2}\frac{(1+\eta_{\tau s}^{2})}{1-\xi_{+}^{2}}=1,
\end{eqnarray}
which indicates
\begin{eqnarray}
\mathcal{N}=\sqrt{\frac{{1-\xi_{+}^{2}}}{1+\eta_{\tau s}^{2}}}.\label{znormal}
\end{eqnarray}

Let us now analyze the effect of $\mathcal{H}_{2}$. For the first two terms
in $\mathcal{H}_{2}$, $\mathcal{H}_{2,1+2}=-\sqrt{3}tq_{x}'\cos\frac{k_{y}}{2}\tau_{z}s_{0}\sigma_{x}
+\sqrt{3}tq_{x}'\sin\frac{k_{y}}{2}\tau_{z}s_{0}\sigma_{y}$,
the corresponding matrix form reads
\begin{eqnarray}
\mathcal{H}_{2,1+2}=\left(
          \begin{array}{ccccccc}
            0 & -\sqrt{3}tq_{x}'\tau_{z}s_{0} & 0 & 0 & 0 & 0 & \cdots \\
            -\sqrt{3}tq_{x}'\tau_{z}s_{0} & 0 & 0 & 0 & 0 & 0 & \cdots \\
            0 & 0 & 0 & -\sqrt{3}tq_{x}'\tau_{z}s_{0} & 0 & 0 & \cdots \\
            0 & 0 & -\sqrt{3}tq_{x}'\tau_{z}s_{0} & 0 & 0 & 0 & \cdots \\
            0 & 0 & 0 & 0 & 0 & -\sqrt{3}tq_{x}'\tau_{z}s_{0} & \cdots \\
            0 & 0 & 0 & 0 & -\sqrt{3}tq_{x}'\tau_{z}s_{0} & 0 & \cdots \\
            \vdots & \vdots & \vdots & \vdots & \vdots & \vdots & \ddots \\
          \end{array}
        \right).
\end{eqnarray}
By projecting $\mathcal{H}_{2,1+2}$ onto $\{|\Psi_{\tau s}\rangle\}$, one finds its contribution
to the boundary Hamiltonian, which reads
\begin{eqnarray}
(\mathcal{H}_{zigzag,1})_{\tau s,\tau's'}&=&\langle\Psi_{\tau s}^{\dag}|\mathcal{H}_{2,1+2}|\Psi_{\tau' s'}\rangle\nonumber\\
&=&-2\sqrt{3}tq_{x}'\tau \eta_{\tau s}\delta_{\tau\tau'}\delta_{ss'}
\mathcal{N}^{2}\sum_{n=0}^{\infty}\xi_{+}^{2n}\nonumber\\
&=&-\frac{2\sqrt{3}t\tau \eta_{\tau s}}{1+\eta_{\tau s}^{2}}q_{x}'\delta_{\tau\tau'}\delta_{ss'}\nonumber\\
&=&\frac{2\sqrt{3}t |\eta_{\tau s}|}{1+\eta_{\tau s}^{2}}sq_{x}'\delta_{\tau\tau'}\delta_{ss'}.
\end{eqnarray}
Above in the last step, we have used the facts $\eta_{\tau s}=-\tau s|\eta_{\tau s}|$ and $\tau^{2}=1$. Also
choosing the basis to be $(|\Psi_{11}\rangle,|\Psi_{1-1}\rangle,|\Psi_{-11}\rangle,|\Psi_{-1-1}\rangle)^{T}$,
then $\mathcal{H}_{zigzag,1}$ can be expressed in terms of the Pauli matrices as
\begin{eqnarray}
\mathcal{H}_{zigzag,1}=\frac{2\sqrt{3}t |\eta_{\tau s}|}{1+\eta_{\tau s}^{2}}q_{x}'\tau_{0}s_{z}.
\end{eqnarray}

For the third term, $\mathcal{H}_{2,3}=-2\sqrt{3}\lambda_{so}q_{x}'\tau_{0}s_{z}\sigma_{z}$, its matrix form reads
\begin{eqnarray}
\mathcal{H}_{2,3}=\tau_{0}\otimes s_{z}\otimes\left(
          \begin{array}{ccccccc}
            2\sqrt{3}\lambda_{so}q_{x}' & 0 & 0 & 0 & 0 & 0 & \cdots \\
            0 & -2\sqrt{3}\lambda_{so} q_{x}'& 0 & 0 & 0 & 0 & \cdots \\
            0 & 0 & 2\sqrt{3}\lambda_{so}q_{x}'& 0 & 0 & 0 & \cdots \\
            0 & 0 & 0 & -2\sqrt{3}\lambda_{so}q_{x}'& 0 & 0 & \cdots \\
            0 & 0 & 0 & 0 & 2\sqrt{3}\lambda_{so}q_{x}'& 0 & \cdots \\
            0 & 0 & 0 & 0 & 0 & -2\sqrt{3}\lambda_{so}q_{x}'& \cdots \\
            \vdots & \vdots & \vdots & \vdots & \vdots & \vdots & \ddots \\
          \end{array}
        \right).
\end{eqnarray}
Its contribution  to the boundary
Hamiltonian can be similarly determined, which takes the form
\begin{eqnarray}
(\mathcal{H}_{zigzag,2})_{\tau s,\tau's'}&=&\langle\Psi_{\tau s}^{\dag}|\mathcal{H}_{2,3}|\Psi_{\tau' s'}\rangle\nonumber\\
&=&2\sqrt{3}\lambda_{so}sq_{x}'(\eta_{\tau s}^{2}-1)
\mathcal{N}^{2}\sum_{n=0}^{\infty}\xi_{+}^{2n}\delta_{\tau\tau'}\delta_{ss'}\nonumber\\
&=&\frac{2\sqrt{3}\lambda_{so}(\eta_{\tau s}^{2}-1)s}{1+\eta_{\tau s}^{2}}q_{x}'\delta_{\tau\tau'}\delta_{ss'}.
\end{eqnarray}
Also in terms of the Pauli matrices, its form can be expressed as
\begin{eqnarray}
\mathcal{H}_{zigzag,2}=\frac{2\sqrt{3}\lambda_{so}(\eta_{\tau s}^{2}-1)}{1+\eta_{\tau s}^{2}}q_{x}'\tau_{0}s_{z}.
\end{eqnarray}
A combination of the two contributions gives the full expression for the linear momentum term
in the boundary Dirac Hamiltonian.  For the chemical potential term, its contribution is also
simply
\begin{eqnarray}
\mathcal{H}_{zigzag,3}=-\mu\tau_{z}s_{0}.
\end{eqnarray}
Let us now analyze the contribution from the last piece, the pairing term $\mathcal{H}_{2,5}=-[\Delta_{0}+2\Delta_{2}(-1-\sqrt{3}q_{x}'\cos\frac{3k_{y}}{2})]\tau_{y}s_{y}\sigma_{0}$.
Its explicit matrix form is
\begin{eqnarray}
\mathcal{H}_{2,5}=\left(
          \begin{array}{ccccc}
            -(\Delta_{0}-2\Delta_{2})\tau_{y}s_{y} & 0 & \sqrt{3}\Delta_{2}q_{x}'\tau_{y}s_{y} & 0 & \cdots \\
            0 & -(\Delta_{0}-2\Delta_{2})\tau_{y}s_{y} & 0 & \sqrt{3}\Delta_{2}q_{x}'\tau_{y}s_{y} & \cdots \\
            \sqrt{3}\Delta_{2}q_{x}'\tau_{y}s_{y} & 0 & -(\Delta_{0}-2\Delta_{2})\tau_{y}s_{y} & 0 & \cdots \\
            0 & \sqrt{3}\Delta_{2}q_{x}'\tau_{y}s_{y} & 0 & -(\Delta_{0}-2\Delta_{2})\tau_{y}s_{y} & \cdots \\
            \vdots & \vdots & \vdots & \vdots & \ddots \\
          \end{array}
        \right).
\end{eqnarray}
Its contribution to the boundary Hamiltonian is
\begin{eqnarray}
(\mathcal{H}_{zigzag,4})_{\tau s,\tau' s'}&=&\langle \Psi_{\tau s}|\mathcal{H}_{2,5}|\Psi_{\tau' s'}\rangle\nonumber\\
&=&-(\Delta_{0}-2\Delta_{2})(\tau_{y})_{\tau\tau'}(s_{y})_{ss'}
+\sqrt{3}\Delta_{2}q_{x}'(\tau_{y})_{\tau\tau'}(s_{y})_{ss'}\mathcal{N}^{2}\sum_{n=1}^{+\infty}2[\psi_{n B}\psi_{(n+1)B}+\psi_{(n+1) A}\psi_{(n+2)A}]\nonumber\\
&=&-(\Delta_{0}-2\Delta_{2})(\tau_{y})_{\tau\tau'}(s_{y})_{ss'}.
\end{eqnarray}
In the last step, we have used the fact that the products $\psi_{n B}\psi_{(n+1)B}$ and $\psi_{n A}\psi_{(n+1)A}$ are
always zero as $\psi_{(2n+1) A}=\psi_{2n B}=0$. In terms of the Pauli matrices,
\begin{eqnarray}
\mathcal{H}_{zigzag,4}=-(\Delta_{0}-2\Delta_{2})\tau_{y}s_{y}.
\end{eqnarray}
Taking all contribution togethers, we reach the final expression for the boundary Hamiltonian for
the upper zigzag edge, which reads
\begin{eqnarray}
\mathcal{H}_{zigzag}=\sum_{i=1}^{4}H_{zigzag,i}=v'
q_{x}'\tau_{0}s_{z}-\mu\tau_{z}s_{0}-(\Delta_{0}-2\Delta_{2})\tau_{y}s_{y},\label{zigzagH}
\end{eqnarray}
where
\begin{eqnarray}
v'=\left[\frac{2\sqrt{3}t |\eta_{\tau s}|+2\sqrt{3}\lambda_{so}(\eta_{\tau s}^{2}-1)}{1+\eta_{\tau s}^{2}}\right].
\end{eqnarray}
When $\lambda_{so}\ll t$, one can do an expansion of $\eta_{\tau s}$ about $\lambda_{so}/t$. Only keeping the leading-order term,
the result is
\begin{eqnarray}
|\eta_{\tau, s}|&=&\frac{4t\lambda_{so}^{2}}{\lambda_{so}(\sqrt{t^{2}(t^{2}+16\lambda_{so}^{2})}-t^{2})}\nonumber\\
&\approx&\frac{4t\lambda_{so}^{2}}{\lambda_{so}(t^{2}+8\lambda_{so}^{2}-t^{2})}\nonumber\\
&=&\frac{t}{2\lambda_{so}}.
\end{eqnarray}
When $\lambda_{so}\ll t$, $|\eta_{\tau, s}|\gg1$, so $\eta_{\tau s}^{2}\pm 1\approx\eta_{\tau s}^{2}$,
one finds
\begin{eqnarray}
v'&=&\frac{2\sqrt{3}t |\eta_{\tau s}|+2\sqrt{3}\lambda_{so}(\eta_{\tau s}^{2}-1)}{1+\eta_{\tau s}^{2}}\nonumber\\
&\approx&\frac{2\sqrt{3}t |\eta_{\tau s}|+2\sqrt{3}\lambda_{so}\eta_{\tau s}^{2}}{\eta_{\tau s}^{2}}\nonumber\\
&\approx& 6\sqrt{3}\lambda_{so}.
\end{eqnarray}
In the limit $\mu=0$, the boundary Hamiltonian reduces to the form of Eq.(9) in the main text.
By comparing the analytical results with the numerical results, we find that the above
low-energy boundary Hamiltonian gives a very accurate description of the physics
on the zigzag edge.

\section{II. Two-dimensional honeycomb-lattice topological insulators in proximity to d-wave superconductors}

In the main text, we have used the isotropic extended s-wave pairing to illustrate the physics.
To show explicitly that the physics does not rely on a specific pairing type, in this section we
consider the d-wave pairing, a pairing type  widely believed to be relevant to high-$T_{c}$ cuprate-based superconductors.  On a honeycomb lattice,
the pairing amplitude of the d-wave pairing follows the pattern $\Delta_{\alpha,ij}=\Delta_{\alpha}\cos 2\theta_{ij}$,
with $\theta_{ij}$ denoting the angle that the bond vector $\bd_{ij}$ is in regard to the $x$ direction.
Accordingly, the Kane-Mele model with d-wave pairing
up to the next-nearest neighbors reads
\begin{eqnarray}
\mathcal{H}_{\rm BdG}(\bk)&=&t(2\cos\frac{\sqrt{3}k_{x}}{2}\cos\frac{k_{y}}{2}+\cos k_{y})\tau_{z}s_{0}\sigma_{x}
-t(2\cos\frac{\sqrt{3}k_{x}}{2}\sin\frac{k_{y}}{2}-\sin k_{y})\tau_{z}s_{0}\sigma_{y}\nonumber\\
&&+2\lambda_{so}(\sin\sqrt{3}k_{x}-2\sin\frac{\sqrt{3}k_{x}}{2}\cos\frac{3k_{y}}{2})\tau_{0}s_{z}\sigma_{z}
-\mu\tau_{z}s_{0}\sigma_{0}\nonumber\\
&&-\Delta_{1}(\cos\frac{\sqrt{3}k_{x}}{2}\cos\frac{k_{y}}{2}-\cos k_{y})\tau_{y}s_{y}\sigma_{x}
+\Delta_{1}(\cos\frac{\sqrt{3}k_{x}}{2}\sin\frac{k_{y}}{2}+\sin k_{y})\tau_{y}s_{y}\sigma_{y}\nonumber\\
&&-2\Delta_{2}(\cos\sqrt{3}k_{x}-\cos\frac{\sqrt{3}k_{x}}{2}\cos\frac{3k_{y}}{2})\tau_{y}s_{y}\sigma_{0}.
\end{eqnarray}
Similar to the extended s-wave pairing, we find that the nearest-neighbor pairing also has
a negligible effect to the helical edge states, and only the next-nearest-neighbor pairing
can open a gap to the boundary Dirac points, as shown in Fig.\ref{dwave}. Therefore, below we also set $\Delta_1=0$
for simplicity. In parallel to the extended s-wave pairing case, we first derive the corresponding low-energy
boundary Hamiltonian for both beard and zigzag edges, and then numerically show the realization of
Majorana Kramers pairs at the sublattice domain walls once the topological criterion is fulfilled.

\begin{figure}[t]
\centering
\includegraphics[scale=0.4]{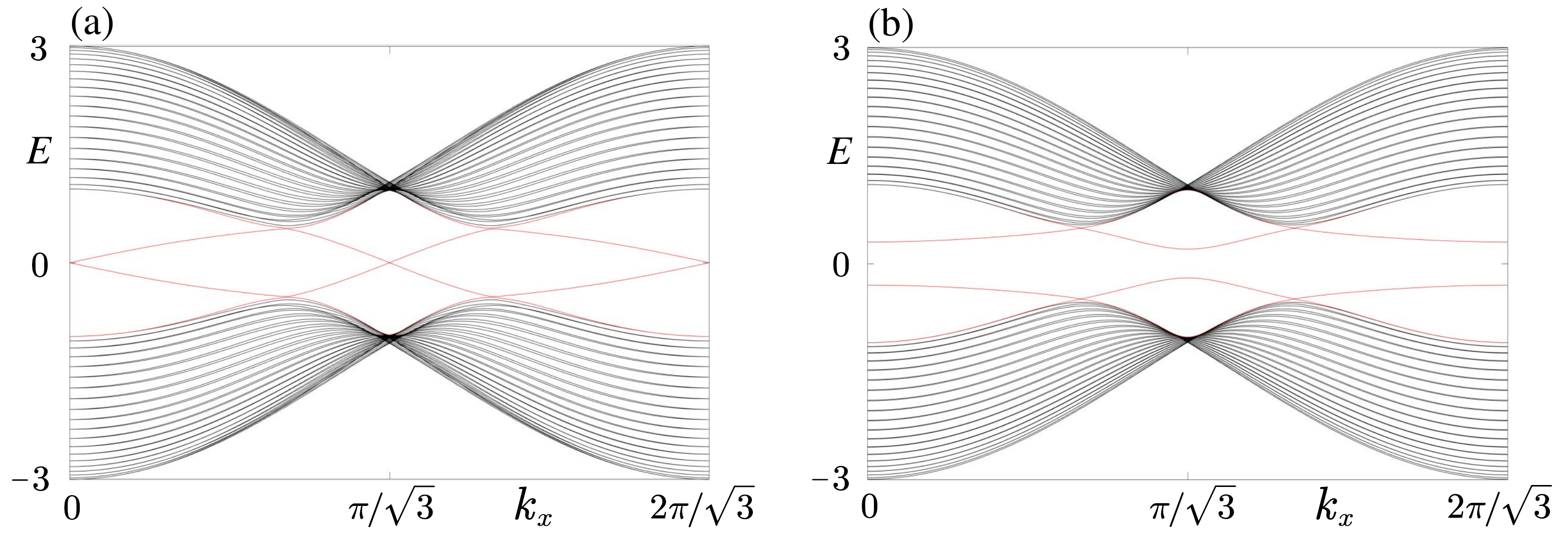}
\caption{(Color online) Energy spectrum for the Kane-Mele model with d-wave pairing.
A cylindrical geometry is considered, with periodic boundary condition in the $x$ direction and open
boundary condition in the $y$ direction. The upper $y$-normal boundary is a beard edge,
and the lower $y$-normal boundary is a zigzag edge. In (a) and (b), $t=1$, $\lambda_{so}=0.1$,
$\mu=0$. In (a),  $\Delta_{1}=0.1$ and $\Delta_{2}=0$, the boundary Dirac points
at time-reversal invariant momentums turn out to be robust against the nearest-neighbor
pairing. In (b),  $\Delta_{1}=0$ and $\Delta_{2}=0.1$, the boundary Dirac points are gapped
by the next-nearest-neighbor pairing. }\label{dwave}
\end{figure}

\subsection{A.  Low-energy boundary Hamiltonian for the beard edge}

Compared to the extended s-wave pairing case, since only the pairing term has been changed, what we need to concern
is just the change of Dirac mass. For the beard edge, as aforementioned the expansion of the Hamiltonian is around
$k_{x}=0$. Also only remaining terms up to the linear order in momentum, the Hamiltonian reads
\begin{eqnarray}
\mathcal{H}_{\rm BdG}(q_{x},k_{y})&=&t(2\cos\frac{k_{y}}{2}+\cos k_{y})\tau_{z}s_{0}\sigma_{x}
-t(2\sin\frac{k_{y}}{2}-\sin k_{y})\tau_{z}s_{0}\sigma_{y}\nonumber\\
&&+2\sqrt{3}\lambda_{so}q_{x}(1-\cos\frac{3k_{y}}{2})\tau_{0}s_{z}\sigma_{z}
-\mu\tau_{z}s_{0}\sigma_{0}\nonumber\\
&&-2\Delta_{2}(1-\cos\frac{3k_{y}}{2})\tau_{y}s_{y}\sigma_{0},
\end{eqnarray}
where the last line corresponds to the pairing term at the neighborhood of $k_{x}=0$.
Focusing on the last line, its matrix form in the underlying basis
$\Psi_{q_{x}}=(c_{1,A,q_{x}},c_{1,B,q_{x}},c_{2,A,q_{x}},c_{2,B,q_{x}},
...,c_{n,A,q_{x}},c_{n,B,q_{x}},...)^{T}$ is
\begin{eqnarray}
\mathcal{H}_{dSC}=\left(
          \begin{array}{ccccc}
            -2\Delta_{2}\tau_{y}s_{y} & 0 & \Delta_{2}\tau_{y}s_{y} & 0 & \cdots \\
            0 & -2\Delta_{2}\tau_{y}s_{y} & 0 & \Delta_{2}\tau_{y}s_{y} & \cdots \\
            \Delta_{2}\tau_{y}s_{y} & 0 & -2\Delta_{2}\tau_{y}s_{y} & 0 & \cdots \\
            0 & \Delta_{2}\tau_{y}s_{y} & 0 & -2\Delta_{2}\tau_{y}s_{y} & \cdots \\
            \vdots & \vdots & \vdots & \vdots & \ddots \\
          \end{array}
        \right).
\end{eqnarray}
Also taking this term as a perturbation and projecting it onto the subspace spanned by the
four eigenvectors $|\Psi_{\tau s}\rangle$, where
\begin{eqnarray}
|\Psi_{\tau s}\rangle=|\tau_{z}=\tau\rangle\otimes|s_{z}=s\rangle\otimes\frac{\sqrt{3}}{2}(1,0,-\frac{1}{2},0,...,
(-\frac{1}{2})^{(n-1)},0,...)^{T},
\end{eqnarray}
one finds that its contribution is
\begin{eqnarray}
(\mathcal{H}_{beard,3})_{\tau s,\tau' s'}&=&\langle \Psi_{\tau s}|\mathcal{H}_{dSC}|\Psi_{\tau' s'}\rangle\nonumber\\
&=& -2\Delta_{2}(\tau_{y})_{\tau\tau'}(s_{y})_{ss'}-\Delta_{2}(\tau_{y})_{\tau\tau'}(s_{y})_{ss'}\nonumber\\
&=&-3\Delta_{2}(\tau_{y})_{\tau\tau'}(s_{y})_{ss'}.
\end{eqnarray}
In terms of the Pauli matrices, its form is
\begin{eqnarray}
\mathcal{H}_{beard,3}=-3\Delta_{2}\tau_{y}s_{y}.
\end{eqnarray}
Replacing the Dirac mass term in Eq.(\ref{beardH}) by the above Dirac mass term, one gets
\begin{eqnarray}
\mathcal{H}_{beard}=vq_{x}\tau_{0}s_{z}-\mu\tau_{z}s_{0}-3\Delta_{2}\tau_{y}s_{y}.
\end{eqnarray}
This is the low-energy boundary Hamiltonian for the beard edge when the pairing is d-wave type.

\subsection{B. Low-energy boundary Hamiltonian for the zigzag edge}

For zigzag edge and d-wave pairing, an expansion of the Hamiltonian around $k_x=\pi/\sqrt{3}$
up to the linear order in momentum gives
\begin{eqnarray}
\mathcal{H}_{\rm BdG}(q_{x}',k_{y})&=&t(-\sqrt{3}q_{x}'\cos\frac{k_{y}}{2}+\cos k_{y})\tau_{z}s_{0}\sigma_{x}
+t(\sqrt{3}q_{x}'\sin\frac{k_{y}}{2}+\sin k_{y})\tau_{z}s_{0}\sigma_{y}\nonumber\\
&&+2\lambda_{so}(-\sqrt{3}q_{x}'-2\cos\frac{3k_{y}}{2})\tau_{0}s_{z}\sigma_{z}
-\mu\tau_{z}s_{0}\sigma_{0}\nonumber\\
&&-2\Delta_{2}(-1+\frac{\sqrt{3}}{2}q_{x}'\cos\frac{3k_{y}}{2})\tau_{y}s_{y}\sigma_{0}.
\end{eqnarray}
where the last line corresponds to the pairing term at the neighborhood of $k_x=\pi/\sqrt{3}$.

In the basis $\Psi_{q_{x}'}=(c_{1,B,q_{x}'},c_{2,A,q_{x}'},c_{2,B,q_{x}'},c_{3,A,q_{x}'},c_{3,B,q_{x}'},
...,c_{n,A,q_{x}'},c_{n,B,q_{x}'},...)^{T}$, the matrix form of the pairing term reads
\begin{eqnarray}
\mathcal{H}_{dSC}=\left(
          \begin{array}{ccccc}
            2\Delta_{2}\tau_{y}s_{y} & 0 & -\frac{\sqrt{3}}{2}\Delta_{2}q_{x}'\tau_{y}s_{y} & 0 & \cdots \\
            0 & 2\Delta_{2}\tau_{y}s_{y} & 0 & -\frac{\sqrt{3}}{2}\Delta_{2}q_{x}'\tau_{y}s_{y} & \cdots \\
            -\frac{\sqrt{3}}{2}\Delta_{2}q_{x}'\tau_{y}s_{y} & 0 & 2\Delta_{2}\tau_{y}s_{y} & 0 & \cdots \\
            0 &  -\frac{\sqrt{3}}{2}\Delta_{2}q_{x}'\tau_{y}s_{y} & 0 & 2\Delta_{2}\tau_{y}s_{y} & \cdots \\
            \vdots & \vdots & \vdots & \vdots & \ddots \\
          \end{array}
        \right).
\end{eqnarray}
Also taking this term as a perturbation and projecting it onto the subspace spanned by the
four eigenvectors $|\Psi_{\tau s}\rangle$, where
\begin{eqnarray}
|\Psi_{\tau s}\rangle=|\tau_{z}=\tau\rangle\otimes|s_{z}=s\rangle\otimes\mathcal{N}(\eta_{\tau s},1,0,0,\xi_{+}\eta_{\tau s},\xi_{+},0,0,\xi_{+}^{2}\eta_{\tau s},\xi_{+}^{2},...)^{T}
\end{eqnarray}
with $\xi$, $\eta_{\tau s}$ and $\mathcal{N}$ given by Eqs.(\ref{zlength}), (\ref{ratio}) and
(\ref{znormal}), respectively, one finds that its contribution is
\begin{eqnarray}
(\mathcal{H}_{zigzag,5})_{\tau s,\tau' s'}&=&\langle \Psi_{\tau s}|\mathcal{H}_{dSC}|\Psi_{\tau' s'}\rangle\nonumber\\
&=&2\Delta_{2}(\tau_{y})_{\tau\tau'}(s_{y})_{ss'}.
\end{eqnarray}
In terms of the Pauli matrices, its form can be expressed as
\begin{eqnarray}
\mathcal{H}_{zigzag,5}=2\Delta_{2}\tau_{y}s_{y}.
\end{eqnarray}
Replacing the Dirac mass term in Eq.(\ref{beardH}) by the above Dirac mass term, one gets
\begin{eqnarray}
\mathcal{H}_{zigzag}=v'q_{x}'\tau_{0}s_{z}-\mu\tau_{z}s_{0}+2\Delta_{2}\tau_{y}s_{y}.
\end{eqnarray}
This is the low-energy boundary Hamiltonian for the zigzag edge when the pairing
is d-wave type.

\begin{figure}[t]
\centering
\includegraphics[scale=0.5]{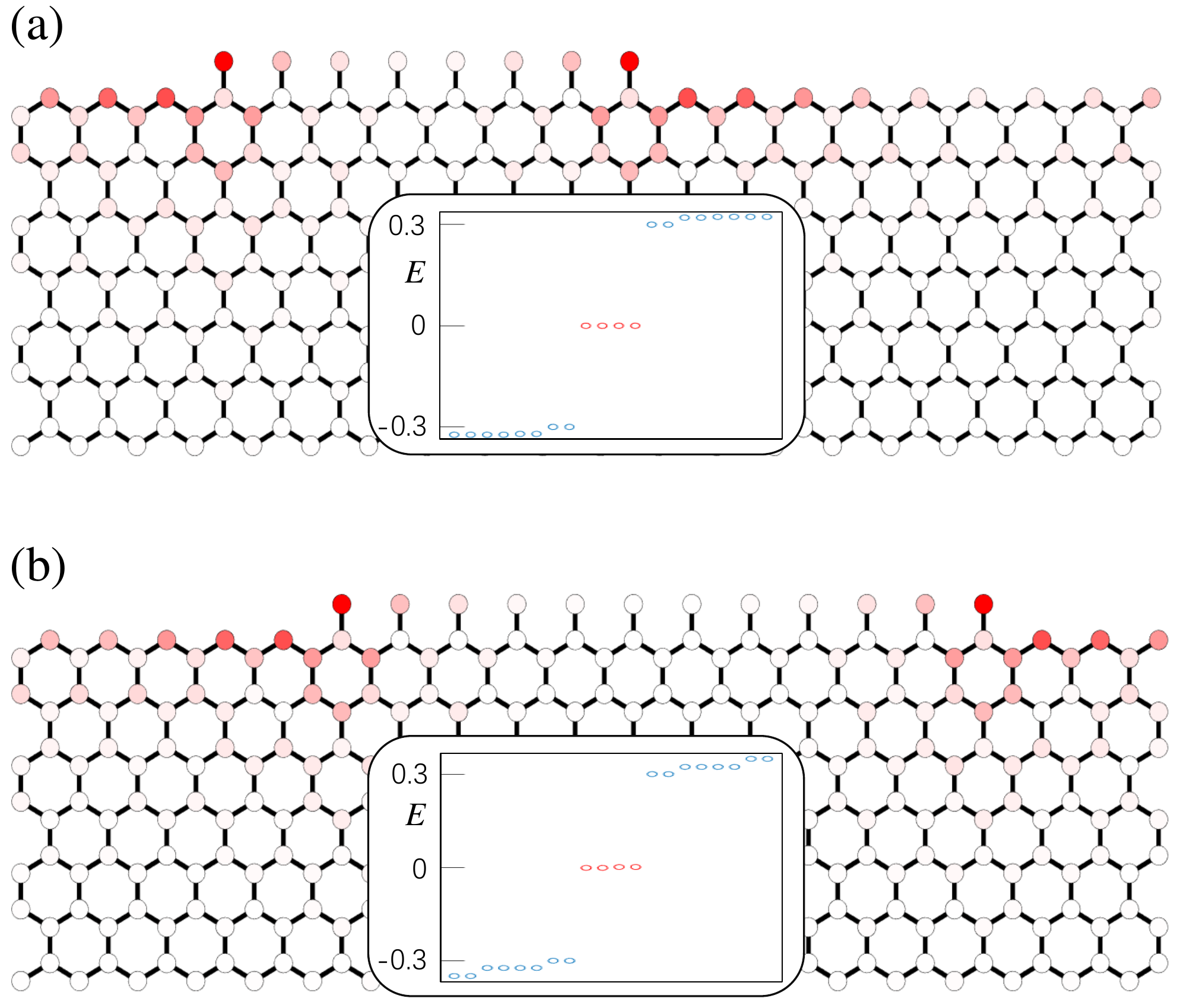}
\caption{(Color online) Majorana Kramers pairs at sublattice domain walls
on the upper boundary. The lattice geometry is
cylindrical  with periodic boundary condition in the $x$ direction
except for the upper most beard-edge part (left and right edges are considered to be
connected in our numerical calculations). Throughout this work,
the same boundary conditions are applied to all similar lattice geometries, below
the boundary conditions will no longer be emphasized when similar lattice geometries
are present.
In (a) and (b), $t=1$, $\lambda_{so}=0.1$, $\mu=0$, $\Delta_{1}=0$, $\Delta_{2}=0.15$,
and the considered lattice geometry and size are shown explicitly. The insets on
top of the underlying lattices correspond to the energy spectra. Here we have only shown
the part of eigenvalues closest to zero energy. The middle four dots in red indicate the
existence of two Majorana Kramers pairs. The shade of red color on the lattice sites
reflect the weight of the probability density ($|\psi(x,y)|^{2}$) of Majorana Kramers pairs.
}\label{dMKP}
\end{figure}

\subsection{C. Majorana Kramers pairs at sublattice domain walls on the upper boundary}

Let us focus on the special case with $\mu=0$.
According to the low-energy boundary Hamiltonian for both beard and zigzag edges, one knows that
if the upper boundary becomes nonuniform and consists of two parts which respectively
terminate at A and B sublattices, the corresponding low-energy boundary Hamiltonian, due to the
further breaking of translation symmetry in the $x$ direction, will become
\begin{eqnarray}
\mathcal{H}=-iv(x)\partial_{x}\tau_{0}s_{z}+m(x)\tau_{y}s_{y},
\end{eqnarray}
where $v(x)=v$, $m(x)=-3\Delta_{2}$ if
the part corresponds to a beard edge, and
$v(x)=v'$, $m(x)=2\Delta_{2}$ if
the part corresponds to a zigzag edge. The velocities in both parts have the same sign, but the
Dirac masses have opposite signs, as a result, the sublattice domain walls correspond to domain walls
of Dirac mass as long as $\Delta_{2}\neq0$. This conclusion suggests that  the sublattice
domain walls will bind Majorana Kramers pairs if the next-nearest-neighboring pairing is finite.
Through numerical calculations, we confirm the appearance of Majorana Kramers pairs at the sublattice
domain walls, as shown in Fig.\ref{dMKP}. The results in this section demonstrate that the
emergence of Majorana zero modes at sublattice domain walls is not restricted
to certain specific pairing type.

\begin{figure}[t]
\centering
\includegraphics[scale=0.5]{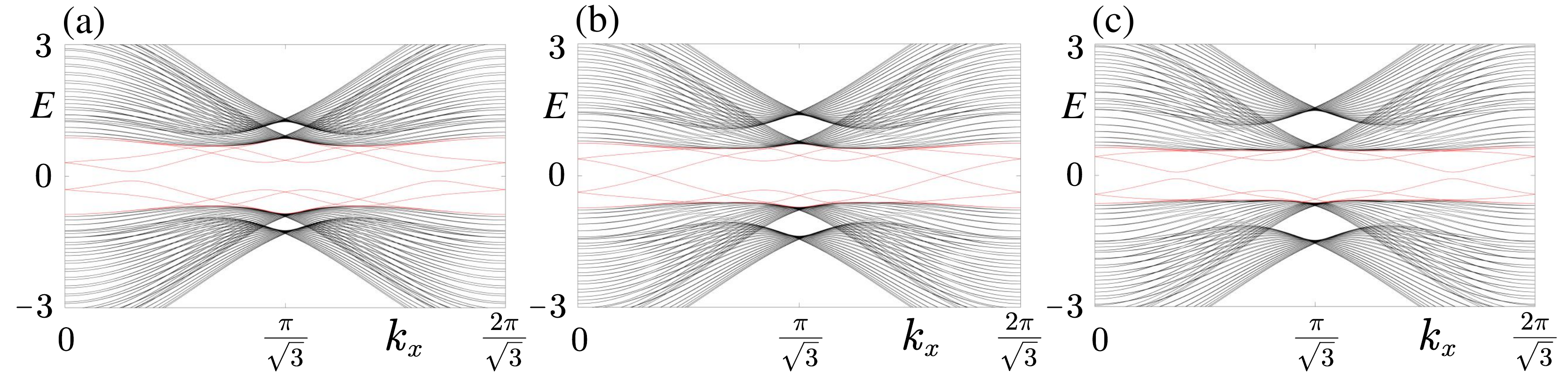}
\caption{(Color online) The evolution of boundary energy gap with respect
to $\mu$. A cylindrical geometry is considered, with periodic boundary condition in the $x$ direction and open
boundary condition in the $y$ direction. The upper $y$-normal boundary is a beard edge,
and the lower $y$-normal boundary is a zigzag edge.
Here the extended s-wave pairing is considered.
In (a)-(c), $t=1$, $\lambda_{so}=0.1$, $\Delta_{0}=\Delta_{2}=0.3$, $\Delta_{1}=0$.
(a) $\mu=0.2$, the mid-gap red lines show the existence of a finite gap in the boundary energy spectrum.
(b) $\mu=0.345$, the boundary energy gap vanishes, corresponding
to the critical point of a boundary topological phase transition.
(c) $\mu=0.45$, a gap is reopened in the boundary energy spectrum. }\label{mugap}
\end{figure}

\section{III. The impact of finite chemical potential on the topological criterion}

Above we have restricted to the $\mu=0$ case for illustration. As the robustness of Majorana
Kramers pairs is protected by non-spatial time-reversal symmetry and particle-hole symmetry and the chemical
potential term does not break these two symmetries, the Majorana Kramers pairs
will remain robust as long as the chemical potential is lower than a critical value.
Before proceeding, it is worth noting that in this work we consider that the superconductivity in
the two-dimensional topological insulator is induced by proximity effect.
Accordingly, the chemical potential is not required to cross the bulk conduction or
valence band to guarantee a metallic normal state to achieve superconductivity.

\begin{figure}[t]
\centering
\includegraphics[scale=0.6]{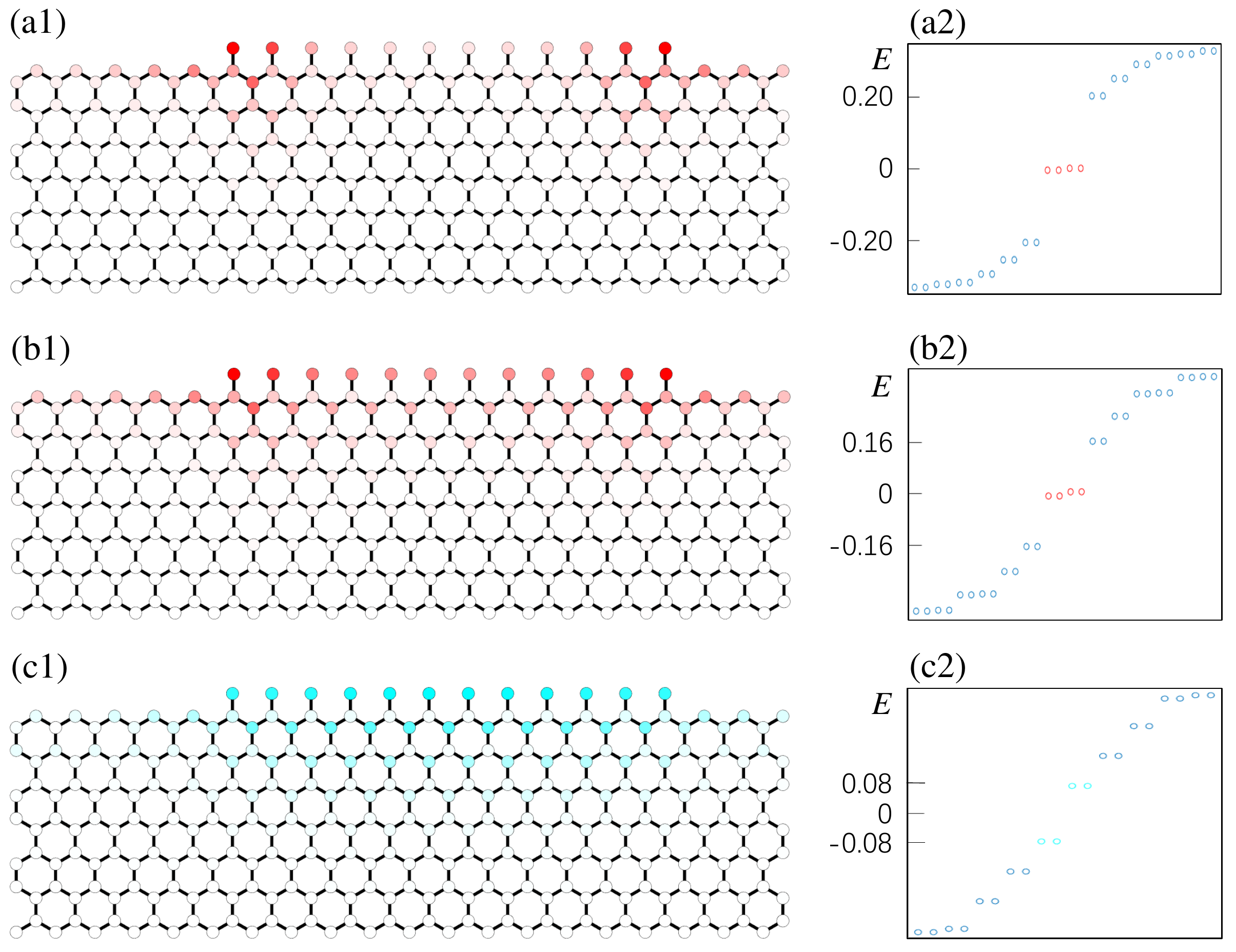}
\caption{(Color online) The effect of chemical potential to Majorana Kramers pairs.
In (a1)-(c2), $t=1$, $\lambda_{so}=0.1$, $\Delta_{0}=\Delta_{2}=0.3$, $\Delta_{1}=0$.
In (a1) and (a2), $\mu=0.1$, the wave functions of Majorana Kramers pairs remain well-localized
so their energy splitting (see red dots in (a2)) induced by the overlap of wave functions
in the considered finite-size system remains small.
In (b1) and (b2), $\mu=0.2$, the increase of $\mu$ reduces the boundary energy gap.
As a result, the localization of the wave functions of Majorana Kramers pairs becomes relatively
poorer and the energy splitting increases due to the finite size of the geometry.
In (c1) and (c2), $\mu=0.4$, no Majorana Kramers pairs are found as the chemical potential
is beyond the critical value. The dots in cyan correspond to the lowest-energy excitations.
One can see from (c1) that their wave functions are quite uniform on the beard edge of
the upper boundary.}\label{muMKP}
\end{figure}

Since the Majorana Kramers pairs have codimension $d_{c}=2$, the critical chemical
potential corresponds to the value at which a boundary topological phase transition
occurs. At the critical point of a boundary topological phase transition, the boundary energy gap vanishes,
while the bulk energy gap can remain open. We can first give an estimate of the critical
value through the low-energy boundary Hamiltonian.  According to Eqs.(\ref{beardH}) and (\ref{zigzagH}),
the Dirac mass changes sign from $k_{x}=0$ to $k_{x}=\pi/\sqrt{3}$
when $|\Delta_{2}|>|\Delta_{0}|/2>0$. It indicates that there exists a node
between $0$ and $\pi/\sqrt{3}$. Furthermore, as the $\Delta_{2}$-term cannot
open a gap at $k_{x}=0$ and the $\Delta_{0}$-term opens an equal gap at $k_{x}=0$
and $k_{x}=\pi/\sqrt{3}$, according to the tight-binding form of the pairing term
in Eq.(\ref{BdGH}), the Dirac mass induced by the on-site and next-nearest-neighbor pairings on the boundary
can be approximated as
\begin{eqnarray}
m(k_{x})\approx -\Delta_{0}+2\Delta_{2}(\cos\sqrt{3}k_{x}-\cos \frac{\sqrt{3}}{2}k_{x}).
\end{eqnarray}
For the parameters considered in the main text, $\Delta_{0}=\Delta_{2}=0.3$,
the node determined by the above formula is located at $k_{x,n}\simeq0.73$.
Since this momentum is closer to $k_{x}=0$ than to $k_{x}=\pi/\sqrt{3}$,
we can focus on the low-energy boundary Hamiltonian for the beard edge shown in
Eq.(\ref{beardH}), whose energy spectrum is
\begin{eqnarray}
E=\pm\sqrt{(3\sqrt{3}\lambda_{so}q_{x}\pm\mu)^{2}+m^{2}(q_{x})}.
\end{eqnarray}
Accordingly, the critical chemical potential is approximately given by
\begin{eqnarray}
\mu_{c}\approx3\sqrt{3}\lambda_{so}k_{x,n}\simeq3.8\lambda_{so}.\label{criticalmu}
\end{eqnarray}
For $\lambda_{so}=0.1$, $\mu_{c}\approx0.38$. In Fig.\ref{mugap},
the evolution of boundary energy gap with respect to $\mu$ is shown.
Fig.\ref{mugap}(b) shows that the precise value of $\mu_{c}$ for the given set
of parameters is $0.345$, quite close to the estimated value. The formula
in Eq.(\ref{criticalmu}) indicates that a stronger spin-orbit coupling admits a larger range of
chemical potential within which the sublattice domain walls can host Majorana Kramers pairs.

In Fig.\ref{muMKP}, the numerical results confirm that the Majorana Kramers pairs are robust against
the increase of chemical potential as long as its value remains lower than the critical value.

\section{IV. Sublattice-sensitive Majorana modes in time-reversal symmetry broken systems}

So far, we have restricted to time-reversal invariant cases. In this section, we consider the
introduction of an in-plane magnetic field to break the time-reversal symmetry. To be specific,
we consider that the magnetic field is applied in the $x$ direction, in parallel to the
domain walls on the upper boundary. The magnetic field will contribute a Zeeman splitting term of
the form $B_{x}\tau_{z}s_{x}\sigma_{0}$ ($g$-factor and Bohr magneton are absorbed in $B_{x}$
for notational simplicity) to the Hamiltonian. Also treating this term as a perturbation, since
this term contains $\sigma_{0}$ in the sublattice subspace,
one can easily find that it will contribute a Dirac mass term of the form $B_{x}\tau_{z}s_{x}$
for both beard and zigzag edges. For generality, let us assume that the Dirac mass
induced by superconductivity on the beard edge is of the form $m_{b}\tau_{y}s_{y}$ and that on
the zigzag edge is of the form $m_{z}\tau_{y}s_{y}$. Taking into account the contribution from
the Zeeman field, the corresponding low-energy boundary Hamiltonians are
\begin{eqnarray}
H_{beard}&=&vq_{x}\tau_{0}s_{z}-\mu\tau_{z}s_{0}+m_{b}\tau_{y}s_{y}+B_{x}\tau_{z}s_{x},\nonumber\\
H_{zigzag}&=&v'q_{x}'\tau_{0}s_{z}-\mu\tau_{z}s_{0}+m_{z}\tau_{y}s_{y}+B_{x}\tau_{z}s_{x}.\nonumber\\
\end{eqnarray}

\begin{figure}[t]
\centering
\includegraphics[scale=0.5]{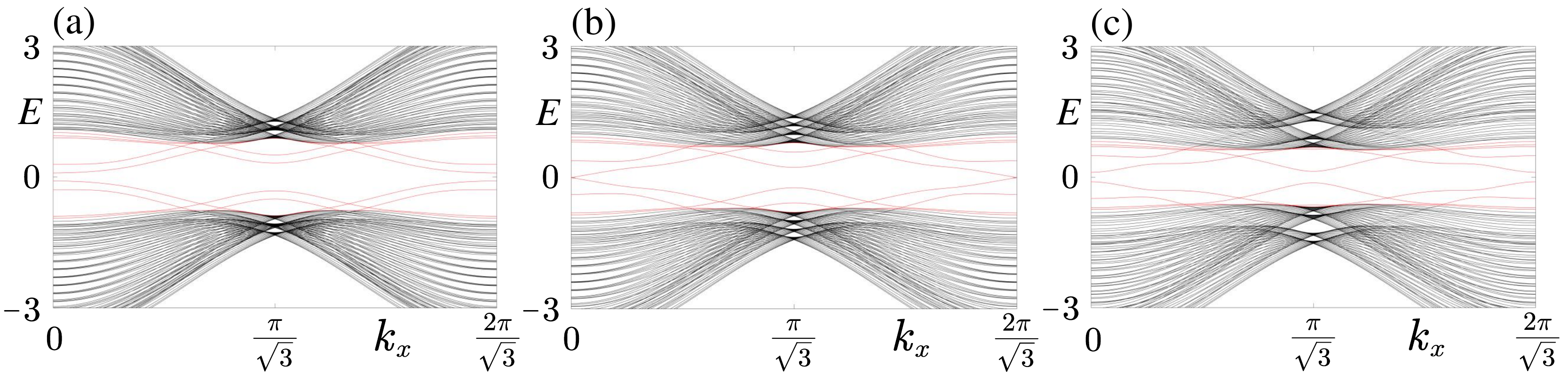}
\caption{(Color online) The evolution of boundary energy gap with respect
to $\mu$. The lattice geometry is cylindrical  with periodic boundary condition in the $x$ direction
and open boundary condition in the $y$ direction. The upper $y$-normal boundary is a beard edge,
and the lower $y$-normal boundary is a zigzag edge. Here the extended s-wave pairing is considered.
In (a)-(c), $t=1$, $\lambda_{so}=0.1$, $\Delta_{0}=0.2$, $\Delta_{2}=0.3$, $\Delta_{1}=0$.
(a) $B_{x}=0.1$, the mid-gap red lines show the existence of a finite gap in the boundary energy spectrum. (b) $B_{x}=0.191$,
the boundary energy gap vanishes, corresponding
to the critical point of a boundary topological phase transition.
(c) $B_{x}=0.3$, a gap is reopened in the boundary energy spectrum.}\label{Zeemangap}
\end{figure}

\begin{figure}[t]
\centering
\includegraphics[scale=0.5]{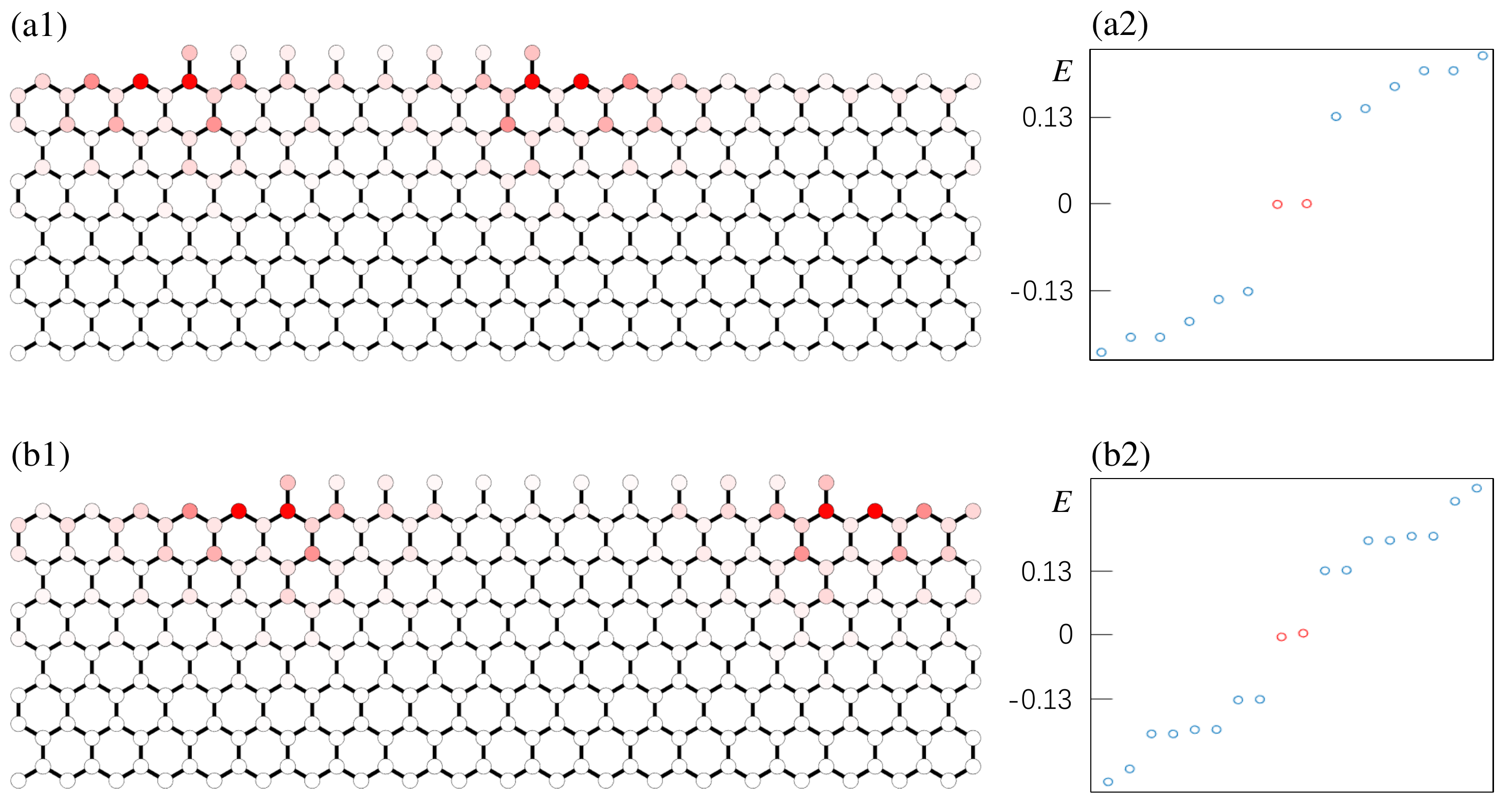}
\caption{(Color online) Majorana zero modes at sublattice domain walls.
The considered lattice geometry and size are shown explicitly. Here the extended s-wave pairing is considered.
In (a1)-(b2), $t=1$, $\lambda_{so}=0.1$, $\mu=0.1$, $\Delta_{0}=0.2$, $\Delta_{2}=0.3$,
$\Delta_{1}=0$, $B_{x}=0.3$. (a1) and (b1) show the distribution of probability density
profiles of Majorana zero modes, and (a2) and (b2) show the corresponding energy spectra, respectively.
Here we have also only shown the part of eigenvalues closest to zero energy. The middle two dots in red in (a2) and
(b2) indicate the
existence of two Majorana zero modes. The shade of red color on the lattice sites in (a1) and
(b1) reflect that the wave functions of Majorana zero modes are  strongly localized around the sublattice
domain walls.}\label{Zeemanmzm}
\end{figure}

Similar to the chemical potential, the Zeeman field will induce a boundary topological phase transition
when it induces a gap closure in the boundary energy spectrum. According to the boundary energy spectrum,
\begin{eqnarray}
E_{beard}(q_{x})&=&\pm\sqrt{v^{2}q_{x}^{2}+\mu^{2}+m_{b}^{2}+B_{x}^{2}\pm2\sqrt{\mu^{2}v^{2}q_{x}^{2}+B_{x}^{2}(\mu^{2}+m_{b}^{2})}},\nonumber\\
E_{zigzag}(q_{x}')&=&\pm\sqrt{v^{'2}q_{x}^{'2}+\mu^{2}+m_{z}^{2}+B_{x}^{2}\pm2\sqrt{\mu^{2}v^{'2}q_{x}^{'2}
+B_{x}^{2}(\mu^{2}+m_{z}^{2})}}.
\end{eqnarray}
For both beard and zigzag edges, the boundary energy gaps will get closed at the time-reversal invariant momentum, i.e., $q_{x}=0$, $q_{x}'=0$.
For the beard edge, the closure of boundary energy gap occurs when $|B_{x}|=B_{b,c}\equiv\sqrt{\mu^{2}+m_{b}^{2}}$.
For the zigzag edge, the condition is similar, that is, $|B_{x}|=B_{z,c}\equiv\sqrt{\mu^{2}+m_{z}^{2}}$.
The critical conditions for the two types of edges indicate that if the Zeeman field is chosen to satisfy
$\text{min}\{B_{b,c},B_{z,c}\}<B_{x}<\text{max}\{B_{b,c},B_{z,c}\}$, the system will enter a new topological
phase on the boundary. For this new topological phase, the Dirac mass of sublattice domain walls will become dominated
by Zeeman field on one side and by superconductivity on the other side. Accordingly, one sublattice
domain wall will change to host a single Majorana zero mode, instead of a Majorana Kramers pair
due to the breaking of time-reversal symmetry.

In Fig.\ref{Zeemangap}, the evolution of boundary energy gap with respect to Zeeman field is shown. One can
readily see that the boundary energy gap gets closed and reopened with the increase of $B_{x}$, in
agreement with the behavior predicted by the low-energy boundary Hamiltonian.
In Fig.\ref{Zeemanmzm}, the numerical results show that each sublattice domain wall hosts one Majorana
zero mode when the topological criterion $\text{min}\{B_{b,c},B_{z,c}\}<B_{x}<\text{max}\{B_{b,c},B_{z,c}\}$
is fulfilled. The results indicate that Majorana zero modes at sublattice domain walls can also be achieved
in time-reversal symmetry breaking systems. Here it is worth noting that if
sublattice-dependent magnetism can be induced on the boundary, Majorana zero modes 
can be realized at the sublattice domain walls even considering a pure on-site s-wave pairing.

\section{V. Tuning the positions of Majorana zero modes by electrically controlling
the local boundary potential}

In the main text as well as in Figs.\ref{dMKP} and \ref{Zeemanmzm},
we have shown that the positions of  Majorana zero modes directly follow the change of the
positions of sublattice domain walls, indicating that if the terminating
sublattices can freely be added or removed, the positions of Majorana zero modes
can be manipulated in a site-by-site way. Apparently, this can benefit
the detection as well as the implementation of braiding Majorana zero modes.
In this section, we show that the positions of Majorana zero modes can also
be tuned  by electrically controlling the local potential
on the boundary, even though the sublattice domain walls are fixed.

\begin{figure}[t]
\centering
\includegraphics[scale=0.6]{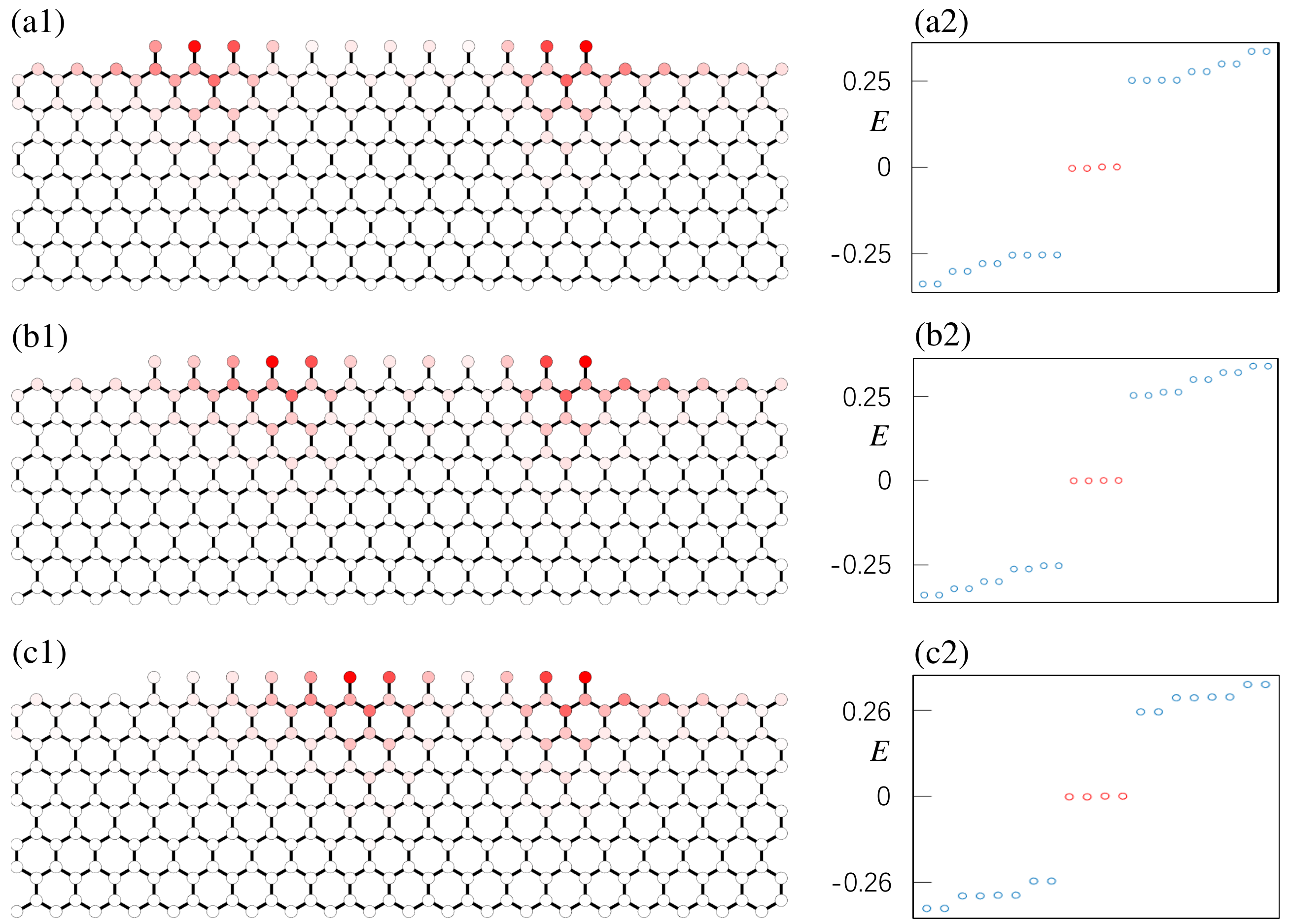}
\caption{(Color online) Manipulating the positions of Majorana zero modes by
electrically controlling the local potential on the uppermost beard edge.
The considered lattice geometry and size are shown explicitly.
Here the extended s-wave pairing is considered.
In (a1)-(c2), $t=1$, $\lambda_{so}=0.1$, $\mu=0$, $\Delta_{0}=\Delta_{2}=0.3$, $\Delta_{1}=0$, $V=2$.
(a1)-(c1) show the distribution of probability density profiles of Majorana zero modes, with
the shade of red color reflecting the weight. (a2)-(c2) are the corresponding energy spectra,
also only the part of eigenvalues closest to zero energy are shown.
On the uppermost beard edge, the lattice sites from left to right are labeled as $1$, $2$, ..., $12$.
In (a1) and (a2), the on-site potential $V$ is only added to the lattice site $1$.
In (b1) and (b2), the on-site potential $V$ is added to lattice sites from $1$ to $3$.
In (c1) and (c2), the on-site potential $V$ is added to lattice sites from $1$ to $5$.
A comparison of the distributions of wave functions of Majorana zero modes in (a1), (b1)
and (c1) clearly shows that the positions of Majorana zero modes can be manipulated by
controlling the local boundary potential.
 }\label{manipulation}
\end{figure}

To show the tunability, we add a coordinate-dependent on-site potential of the form
$\sum_{i}V_{i}c_{i}^{\dag}c_{i}$ to the Hamiltonian, and $V_{i}$ is
chosen to be a nonzero constant only at the neighborhood of the sublattice domain walls.
The considered lattice geometry is shown explicitly in Fig.\ref{manipulation}.
For the convenience of discussion, let us label the lattice sites on the uppermost beard edge
from left to right as $1$, $2$,...,$12$. In Figs.\ref{manipulation}(a1) and (a2),
the on-site potential is only added to site $1$. From the shade of red color on the lattice
sites, it is readily found that the site having the highest weight of the probability density
of Majorana zero modes becomes site $2$. In Figs.\ref{manipulation}(b1) and (b2),
the on-site potential is added to sites from $1$ to $3$. Also from the
shade of red color on the lattice sites, it is readily found that
the site having the highest weight is now shifted from site $2$ to site
$4$. In Figs.\ref{manipulation}(c1) and (c2), the on-site potential is added to sites from $1$ to $5$.
It is readily found that the site having the highest weight changes to site $6$.
The results demonstrate explicitly that the positions of Majorana zero modes
can be manipulated site-by-site by controlling the local boundary potential.

\end{widetext}

\end{document}